\documentstyle[emulateapj,psfig,apjfonts]{article}

\makeatletter

\newenvironment{inlinefigure}{%
\def\@captype{figure}%
\noindent\begin{minipage}{0.999\linewidth}\begin{center}}
{\end{center}\end{minipage}\smallskip}
\makeatother

\def\gs{\mathrel{\raise0.35ex\hbox{$\scriptstyle >$}\kern-0.6em 
\lower0.40ex\hbox{{$\scriptstyle \sim$}}}}
\def\ls{\mathrel{\raise0.35ex\hbox{$\scriptstyle <$}\kern-0.6em 
\lower0.40ex\hbox{{$\scriptstyle \sim$}}}}
\newcommand{\ltaraw}{$\; \buildrel < \over \sim \;$}
\newcommand{\lta}{\lower.5ex\hbox{\ltaraw}}
\newcommand{\gtaraw}{$\; \buildrel > \over \sim \;$}
\newcommand{\gta}{\lower.5ex\hbox{\gtaraw}}

\newcommand{\lsun}{{\rm\,L_\odot}}

\newcommand{\mum}{$\,\mu$m}

\newcommand{\ratio}{${\cal R}(60,100)$}
\newcommand{\bivar}{$\Phi$(${\cal L}$, ${\cal C}$)}
\newcommand{\lfir}{L$_{\rm FIR}$}
\newcommand{\ltir}{L$_{\rm TIR}$}

 \def\itm#1 {\vskip10pt \noindent \square\ {\bf #1} }
  \def\square {\hbox{\vrule width5pt height5pt}}

  \def\cle      {{$_ <\atop{^\sim}$}}
  \def\cge      {{$_ >\atop{^\sim}$}}

  \def\deg      {{\ifmmode^\circ\else$^\circ$\fi} } 
 \def\arcm  {{\ifmmode {'\ }\else$'     $\fi} } 
 \def\arcs  {{\ifmmode{''\ }\else$''    $\fi} } 

\lefthead{Chapman et al.}
\righthead{The bi-variate distribution of IRAS galaxies}

\begin{document}

\title{The bi-variate luminosity-color distribution of
IRAS galaxies, and implications for the high redshift Universe}

\author{S.\,C.\ Chapman,$\!$\altaffilmark{1} 
	G.\ Helou,$\!$\altaffilmark{1}
	G.\,F.\ Lewis,$\!$\altaffilmark{2}
	D.\,A.\ Dale$\!$\altaffilmark{3}
}
\affil{California Institute of Technology,
Pasadena, CA 91125,~~U.S.A.}
\affil{Anglo-Australian Observatory, P.O. Box 296, Epping,
        NSW 1710, Australia}
\affil{Department of Physics \& Astronomy University of Wyoming Larami
e, WY 82071}



\begin{abstract}
We present a characterization of
the local luminosity-color, bi-variate distribution
of IRAS galaxies from the 1.2\,Jy sample, selected at 60\mum.
The \ratio\ infrared color is used
as the best single parameter description of the IR spectral energy
distribution of galaxies.
We derive an analytical form of the distribution and use it
to constrain the effect of the IR color distribution on 
evolution models for high redshift, farIR luminous galaxies.
Our adopted evolution retains the locally observed correlation between
luminosity and color, such that the larger characteristic luminosities
at higher redshift have a warmer characteristic color. The
width of the color distribution at a given luminosity 
remains constant for all redshifts.
We demonstrate that there is the potential for both hotter and colder sources 
to be missed in cosmological surveys. 
An evolving bi-variate luminosity function coupled
with the cold source bias of 
sub-mm selected surveys suggests the existence of a
large population of cold sources appearing in such surveys. 
Likewise, a hot source bias for most SIRTF wavebands together with a
bi-variate model suggests an excess of hot sources being selected.
We test the evolutionary form against available data for higher redshift,
farIR galaxies. The data
do not reveal evidence for any strong evolution in the characteristic
luminosity-color distribution
as a function of redshift over $0<z<1$. However, there is
marginal evidence for a broadening of the color distribution
at higher redshifts, consistent with our locally characterized
trend of a broadening in the IR color distribution at the highest luminosities.

\end{abstract}

\keywords{
galaxies: evolution --- galaxies: formation --- sub-mm: galaxies --- radio:
galaxies --- IR: galaxies}


%

\section{Introduction}\label{introduction}

The local infrared-luminous galaxies detected by the {\it IRAS} satellite
exhibit a vast array of source properties.
Infrared color has typically been used to attempt to parametrize the
IRAS population (Soifer \& Neugebauer 1991).
Dale et al.\ (2001) demonstrated that the 
S$_{\rm 60 \mu m}$/S$_{\rm 100 \mu m}$ flux ratio or color 
(hereafter referred to as \ratio) provides the
best single parameter characterization of the {\it IRAS} galaxy
population, in addition to luminosity.
While a complicated array of dust properties 
contribute to the spectral energy distribution 
(SED) of each galaxy, studies of {\it IRAS} galaxies have typically reduced 
the description to a best fit single dust temperature, T$_{\rm d}$,
with a one-to-one mapping to \ratio.
Indeed, 
changing the dust temperature has been demonstrated to have a significantly
larger effect on the galaxy SED than dust emissivity, mid-IR spectral
index, or cosmology (Blain et al.\ 2002). The inferred luminosity
of an infrared galaxy for a fixed observed mid-IR flux density increases by a
factor of 10 if the dust temperature is doubled.

It has been demonstrated that low-redshift {\it IRAS} galaxies exhibit
slowly varying correlations of
\ratio\ with luminosity 
(Dale et al.\ 2001; Dunne et al.\ 2000; Andreani \& Franceschini 1996).
However, over a
large spread of luminosities the \ratio\ ratio of infrared galaxies
does change systematically. Fitting single dust temperature (T$_{\rm d}$) 
models to \ratio, we find 
$\sim$20\,K for low-redshift spirals (Reach et al.\ 1995; Alton et al.\ 2000;
Dunne \& Eales 2001) and 30\,K--60\,K for the high luminosity objects typically
detected by {\it IRAS} (Soifer \& Neugebauer 1991, Stanford et al.\ 2000).
High redshift, hyper-luminous galaxies can show dust temperatures  
of up to 110\,K  (e.g., Lewis et al.\ 1998), 
implying a continuation in the luminosity-T$_{\rm d}$
relation out to the higher luminosities characteristic
of the distant Universe. 

However, while a statistical relation exists between the \ratio\ and
IR luminosity, the distribution is broad. We find 
in substantial numbers both extremely luminous, yet cold galaxies,
as well as low luminosity, hot galaxies. 
One surprisingly cold and luminous galaxy, Arp 302 (or UGC\,9618/NGC\,5051), has
been identified from the IRAS bright galaxy sample (BGS) with
S$_{\rm 60 \mu m}$=6.8\,Jy,  S$_{\rm 100 \mu m}$=15.3\,Jy, and  
\lfir=3.89$\times10^{11}\lsun$.
This is the system with largest deviation from the median \ratio\ 
for its luminosity.
Lo, Gao \& Guendl (1997) suggest this galaxy is the 
most massive known (in terms of CO gas mass).
By contrast, galaxies with very hot IR color, 
and without obvious AGN contributions, 
have been identified from the faintest \lfir\ sources detected by IRAS.
NGC1377, NGC4491 and IRAS1953 with \ratio$\sim$1 and  
\lfir$\sim10^9\lsun$ are 
as hot or hotter than the 
Ultra-Luminous Infrared Galaxy (ULIG), Arp220, but with 10$^{-3}$ the 
FIR luminosity (Roussel et al.~2003).

The possible 
importance of cold, luminous galaxies to farIR and sub-mm surveys
has been pointed out by Eales et al.\ (1999, 2000). This has recently
been highlighted observationally by Chapman et al.~(2002a) who 
demonstrated that cold and luminous sources exist at higher redshift, 
identifying two {\it ISO}-FIRBACK sources
(FB1-40 and FB1-64) with
\ltir $>10^{12}$ L$\odot$ with $z$\cle 1 galaxies, and finding best fit
single temperature grey-bodies of 26\,K and 31\,K respectively
(dust emissivity $\beta=1.6$).
The local cold source, Arp~302, approximately matches the color and luminosity 
of FB1-40. 
However, FB1-40 and FB1-64 were discovered from a random sampling of the
sub-mm luminous FIRBACK sources, and they cannot be considered
as an insignificant portion of the high-$z$ ULIG population.

It is therefore a concern that studies of the evolving galaxy populations
typically assume a small range of 
template galaxy spectral energy distributions (SEDs), sampling only the 
monotonic relation of dust temperature to luminosity
in choosing the SED templates (Blain et al.\ 1999a,b, 
Malkan \& Stecker 2000, Rowan-Robinson 2001, 
Chary \& Elbaz 2001, Chapman et al.\ 2002b, Franceschini et al.\ 2002).
Modeling efforts
have in effect tied the dust temperature directly to the farIR luminosity,
ignoring the T$_{\rm d}$ distribution for each luminosity class.
This in part may be a result of the fact that
the \ratio\ distribution
for local {\it IRAS} galaxies has never been carefully studied, 
and no analytical 
form for the distribution has been presented.

Importantly, an evolving 
distribution bi-variate in luminosity and color, \bivar,
consistent with the broad distribution observed locally should 
subsume a non-negligible fraction of both cold, luminous galaxies and
hot, faint galaxies. Surveys which select objects at either the
cold Raleigh-Jeans tail of the dust SED, or the hot Wien tail, will
preferentially detect appropriately cold or hot 
objects for a given luminosity class, if they exist in non-negligible
numbers.
For instance, the cold source bias of farIR (e.g., 170\mum) and
submillimeter selected surveys will result in any existing 
population of cold sources being over-represented.

At the highest redshifts, the poorly understood submillimeter (sub-mm)
population is thought to dominate the most luminous IR galaxies.
Since the discovery of high redshift sub-mm sources
(Smail, Ivison \& Blain 1997), there has been an ongoing debate
about their nature and their dust properties.
In the absence of redshifts, it is difficult to understand whether they
represent hot sources at very high redshifts, or colder sources at more
modest redshifts (e.g.~Eales et al.~1999,2000), 
similar to FB1-40 and FB1-64 (Chapman et al.\ 2002a).
Chapman et al.\ (2003a) have measured spectroscopic redshifts for radio
identified sub-mm galaxies, claiming a typical dust temperature
of $\sim$40\,K by assuming the empirical relation between the far-IR and radio
observed locally (e.g., Helou et al., 1985).
However, without additional SED measurements, 
it is unclear what distribution in dust properties for the sub-mm
galaxies remains consistent with their currently measured
properties.

In this paper, we characterize 
the local luminosity-color, bi-variate distribution, \bivar,
of IRAS galaxies from the 1.2\,Jy sample, selected at 60\mum.
We represent the infrared luminosity with \ltir,
and the \ratio\ IR-color is used
as the best single parameter description of the IR SED of galaxies.
\ltir\ is defined as in Dale et al.\ (2001), integrating over the
SED from 3-1100\mum. These authors define a bolometric conversion between
the more typical FIR luminosity (e.g., Helou et al., 1985) as
$$\log({\rm TIR/FIR}) = a_0 + a_1x + a_2x^2 + a_3x^3 + a_4x^4$$
where $x = \log{{f_\nu(60\mu m)}\over{f_\nu(100\mu m)}}$ 
and [$a$($z$=0] = [0.2738, -0.0282, 0.7281, 0.6208, 0.9118].
The first part of the paper is directed at the derivation of
an analytical form for the local IRAS color distribution.
In order to constrain the effect of the IR color distribution on our
understanding of high redshift farIR luminous galaxies, 
we evolve the bi-variate distribution according to luminosity 
evolution prescriptions used in the literature.
For convenience, we use the terms `color' (meaning 60/100 micron flux
ratio) and `temperature' interchangeably, even though we do not
collapse the SED description to a single temperature blackbody.
The color is taken to be a diagnostic of the typical heating conditions in the
ISM of a galaxy, and therefore would be indicative of a characteristic
dust temperature.
Section~2 describes our sample used to construct \bivar.
Section~3 provides a statistical analysis of the 1.2\,Jy sample,
and provides a best fit analytical form to \bivar.
Section~4 explores the evolutionary behavior of \bivar, while
Section~5 compares the evolving \bivar\ to existing data sets at 
higher redshift.
 





\section{Sample selection}

Our starting point for this study is the S$_{\rm 60 \mu m}>1.2$\,Jy 
sample of galaxies
(Fisher et al.~1995). We plot the \ratio\ distribution of these galaxies
as a function of both \lfir\ and \ltir\ in Fig.~1. \lfir\ is calculated
directly from the 60\mum\ and 100\mum\ flux densities and the redshift
as described in Helou et al.\ (1985).
The \ltir\ parameter represents a bolometric correction for the flux
from 3 to 1100 microns, and represents a larger correction for sources
with cooler \ratio\ colors, as seen in the comparison of the two panels
of Fig.~1.  The analytical expression mapping
the average FIR luminosity to TIR luminosity from Dale et al.\ (2001)
was reproduced in the previous section.

While studying this sample, we noticed a significant tail of very cold and
luminous galaxies, departing bi-modally from the main distribution (see
Fig.~1). Upon closer inspection of the 60 and 100\mum\ IRAS fluxes using
the {\bf XSCANPI} application provided by the Infrared Processing and 
Analysis Center (IPAC),
we found that almost 100\% of these sources have spurious measurements.
Firstly, the cirrus contamination
can be large, affecting the 100\mum\ more than 60\mum\ and leading to an
apparently small 60\mum/100\mum\ ratio.
Secondly, the 60\mum\ flux measurements found for some of the sources 
from XSCANPI were  
less than 1.2\,Jy, and should not be included in our catalog.

%
%
\begin{inlinefigure}
\psfig{figure=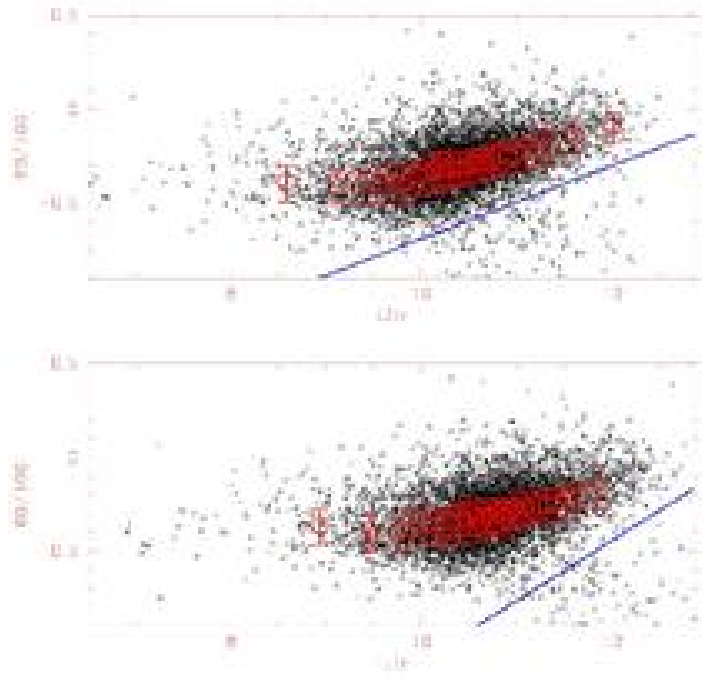,angle=0,width=3.5in}
\vspace{6pt}
\figurenum{1}
\caption{
The \lfir\ and \ltir\ distributions in \ratio. 
Spurious cold, luminous sources have been
removed from the sample, as indicated here by solid lines delimiting the
region of spurious sources. 
The medians in equal number bins are plotted as large circles, 
with the interquartile range
shown in error bars.}
\label{fig1}
\addtolength{\baselineskip}{10pt}
\end{inlinefigure}

We flagged all such sources in our sample 
and removed them from subsequent consideration.
To ensure that this effect did not significantly contaminate the main
distribution, we randomly selected 200 of the sources with \lfir $>10^{11}$
for XSCANPI analysis, finding them to have correct flux estimates to within
5\% of the Fisher et al.~(1995) values.
We then derive the TIR luminosity function with the revised
catalog of 1.2Jy sources. We adopt an accessible volume technique as
described in (Avni \& Bahcall 1980)
Our general luminosity function is represented  as

\begin{equation}
\Phi(L)\Delta L = \sum_i \frac{1}{V_i}    
\end{equation}

with $\Phi(L)\Delta L$ as the number density of sources (Mpc$^{-3}$)
in the luminosity range $L$ to $L+\Delta L$. 
The accessible volume, $V_i$, represents the
{\em i}th source in the sample, the
maximum volume in which the object could be located and still be detected
in the {\em IRAS} 1.2\,Jy 60\mum\ catalog.
The sum is then over all sources within the luminosity range.
We then map sources to their TIR luminosity using the above definition.
Our constructed TIR luminosity function is therefore directly related to the
60\mum\ luminosity function for this sample.

Figure~1 also shows the median \ratio\ values and the interquartile range,
revealing the dust temperature to luminosity relation. 
In the subsequent sections, we shall study and characterize this relation.


%
\section{Results}\label{results}

\subsection{Statistics}\label{X}

In choosing a luminosity variable in \bivar\ to describe the distribution
of galaxies, it is desirable to minimize the dependencies between 
${\cal L}$ and ${\cal C}$.
\ltir\ has been shown to provide less dependency than \lfir\
on IRAS colors (Dale et al.~2001).
Similarly, \ratio\ has been demonstrated to parametrize the variation in
mid--farIR properties better than any other combination of IRAS bands
(Dale et al.~2001).

%
%
\begin{inlinefigure}
\psfig{figure=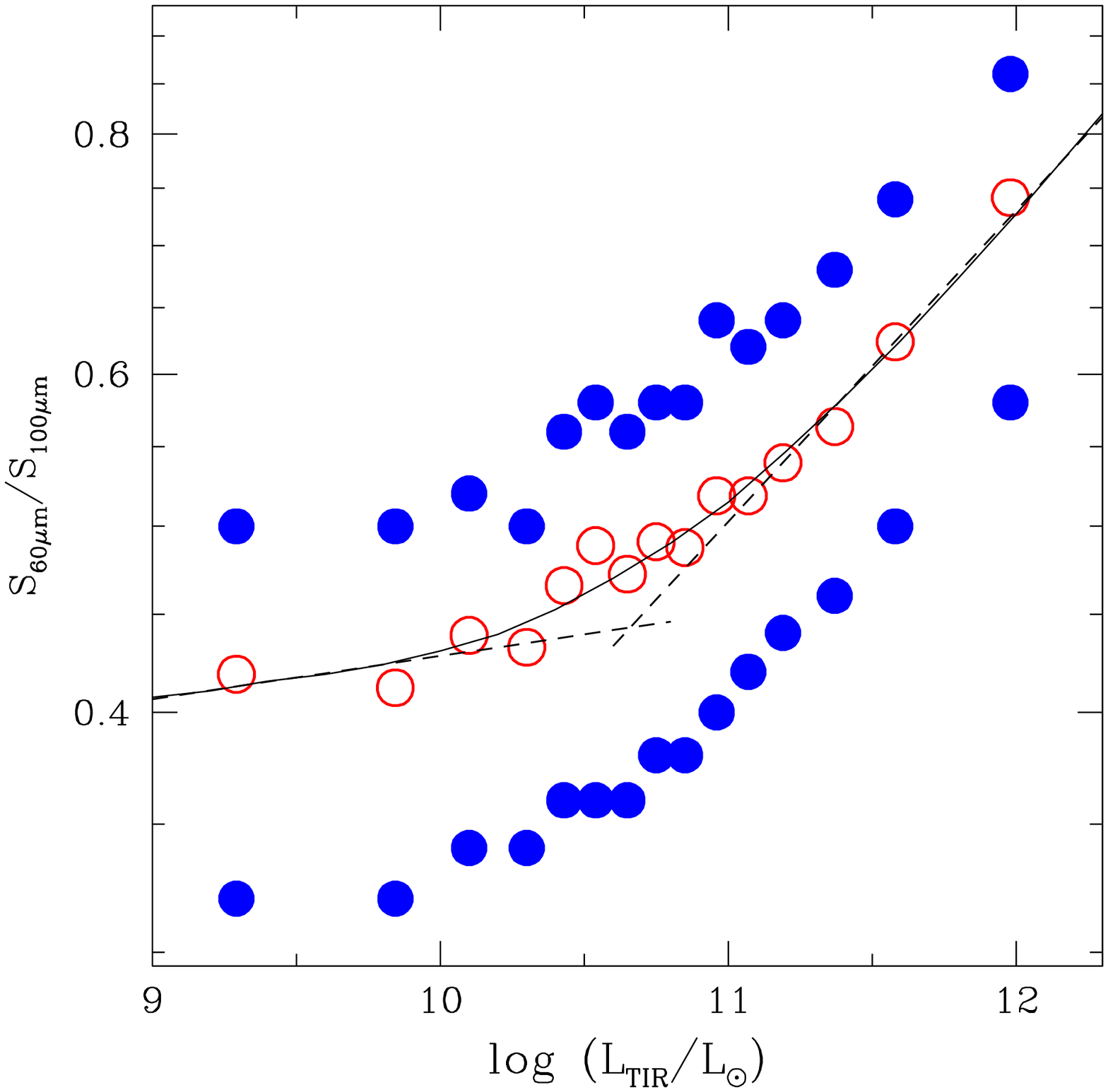,angle=0,width=3.5in}
\psfig{figure=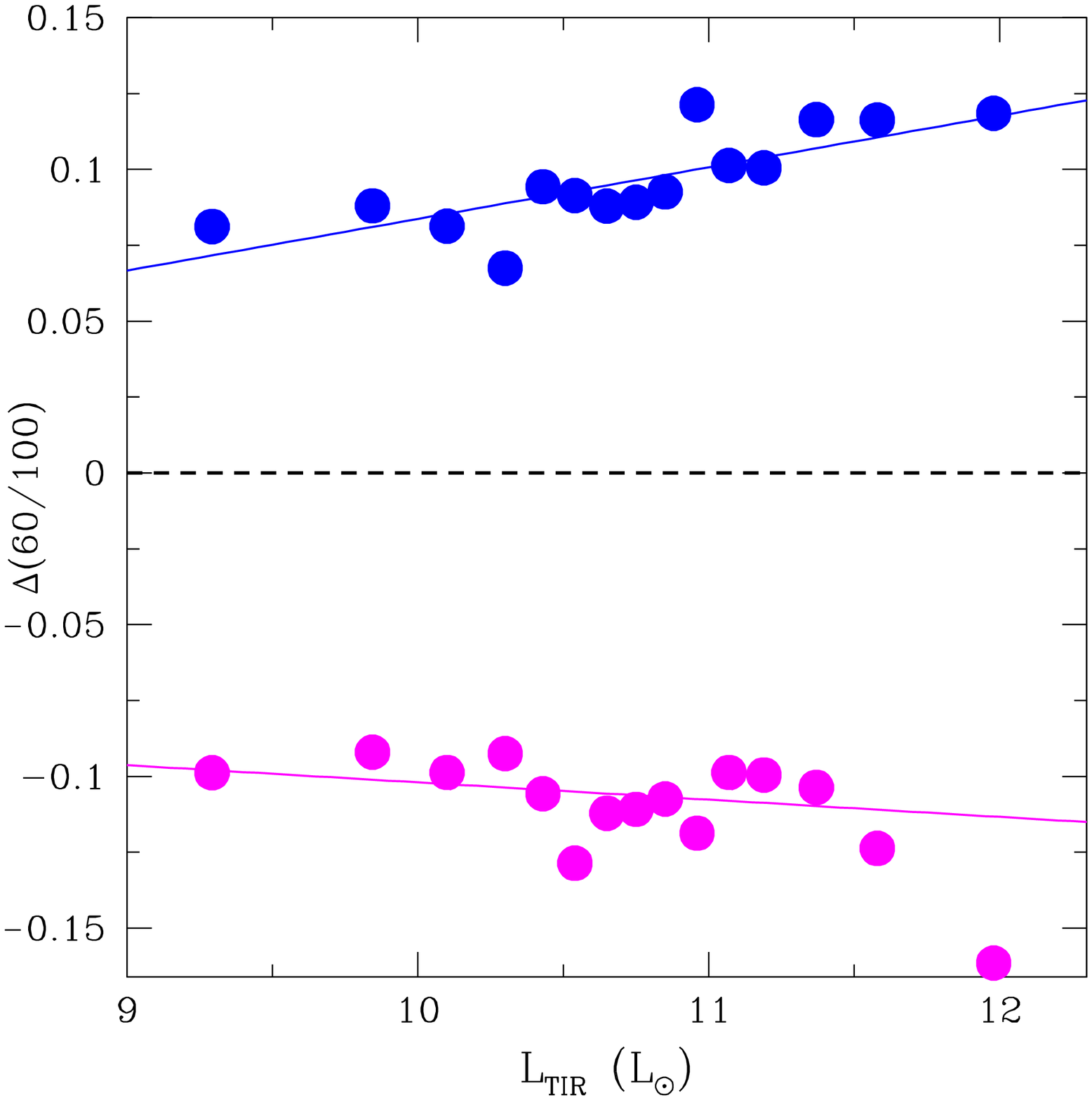,angle=0,width=3.5in}
\vspace{6pt}
\figurenum{2}
\caption{
{\bf Upper Panel:} The 1st quartile, median, and
last quartile for the 1.2\,Jy sample with a log scaling in \ratio,
suggesting a roughly constant width and shape in log space.
{\bf Lower Panel:} The same statistics with the median subtracted,
shown in a linear scaling emphasizing
the small linear increase in width with increasing luminosity.
The width is constant in log(\ratio).}
\label{fig2}
\addtolength{\baselineskip}{10pt}
\end{inlinefigure}

Having chosen our variable representation of \bivar, we can proceed to 
quantify the relation for the S$_{60}$>1.2\,Jy IRAS sample and 
explore the implications.
We begin by calculating the median and first/last quartile statistics for the
1.2\,Jy sample, corrected for spurious cold, luminous sources as described
in section~2.
The analytic relation in the distribution median is well fit by a dual
power law, as shown in figure 2:  
$$\log{\cal R}(60/100) = 0.162 \times \log(L_{\rm TIR/}L_\odot) - 2.080,\ L_{\rm TIR}>5\times10^{10}\,L_\odot$$
$$\log{\cal R}(60/100) = 0.022 \times \log(L_{\rm TIR}/L_\odot) - 0.593,\ L_{\rm TIR}<5\times10^{10}\,L_\odot.$$

We express this composite function with a smooth transition as:
$${\cal R}{(60/100)} = {\cal C_*} \times (1 + {{\cal L_*}\over{L_{\rm TIR}}})^{-\delta} \times (1 + {{L_{\rm TIR}}\over{\cal L_*}})^{\gamma}$$
$${\rm with,} \gamma = 0.16,\, \delta = 0.02,\, {\cal C_*} = 0.45,\,
	{\cal L_*} = 5.0\times10^{10} L_\odot$$

The statistics and the fitted relation are shown in Fig.~2a. We also fit the
width of the inter-quartile statistics after removing the median 
(Fig.~2b). 
The width of the distribution described in terms of this statistic is
essentially constant in $\log$(\ratio) as a function of \ltir, 
therefore showing a slight broadening in \ratio\ as shown in Fig.~2b.
At the highest luminosities there is some evidence for significant
broadening of even the $\log$(\ratio), although the statistics are
poor and there remains a concern that all spurious sources have not
been removed from the sample (section 2).

These equations provide a simple, first order description of the \bivar\ 
distribution of IRAS galaxies from which to gage a more comprehensive
analysis. This expression should be suitable for many applications and
evolutionary extrapolations.

\subsection{Fitting the \bivar\ distribution}\label{X}

We now consider the detailed distributions in \ratio\ for the
1.2\,Jy sample. Histograms of the distribution 
from each luminosity bin are plotted in 
Fig.~3. For clarity, a fixed constant is introduced to offset each
class of \ltir.
With no apriori assumptions about the \ratio\ distribution of the
IRAS sources, we begin by testing whether the population might be well
represented by a gaussian in the \ratio\ color.
A gaussian in both linear \ratio\ and log\ratio\ 
was first fit to each bin of 380 sources. The functional form
$$C + B*x + D* exp[-1/2\times ({{x-A}\over{E}})^2],$$
where $x$ represents \ratio, was fit to each histogram of 380 sources.
An equal weighting of the points in the histogram was assumed in the fitting.
The residuals were compared, with the $\chi^2$ being a factor $\sim$2 
smaller for
the gaussian in log (or {\it log normal} distribution), suggesting
the log normal as the more appropriate representation.

%
%
\begin{inlinefigure}
\psfig{figure=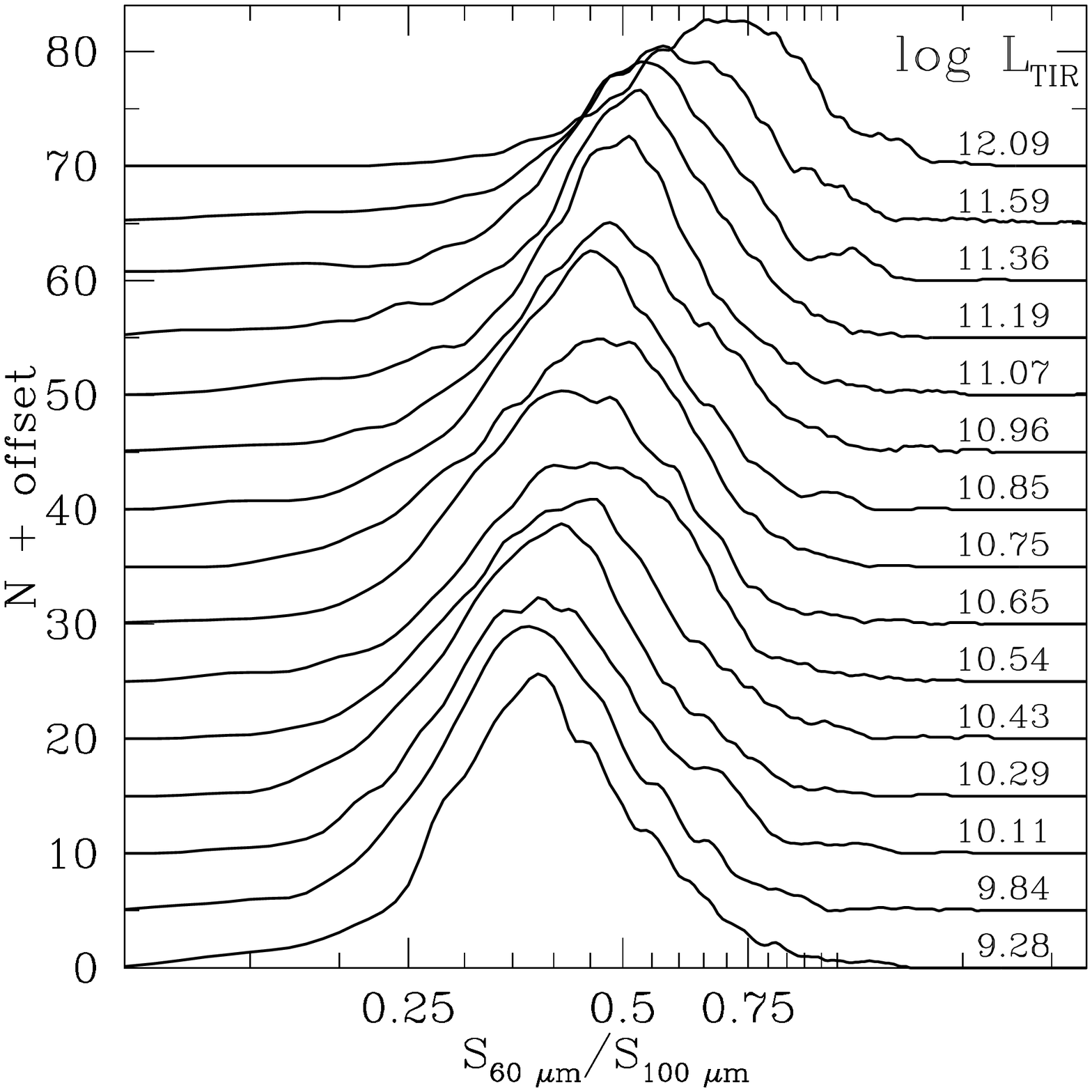,angle=0,width=3.5in}
\vspace{6pt}
\figurenum{3}
\caption{
The \bivar\ distribution in log IR-color, shown as a smoothed histogram.
The effective histogram
bin size is d\,\ratio=0.05. 
Histograms of \ltir\ are shown offset by a fixed constant for clarity, and
labeled on the right in $\lsun$.
}
\label{fig3}
\addtolength{\baselineskip}{10pt}
\end{inlinefigure}

%
%
\begin{inlinefigure}
\psfig{figure=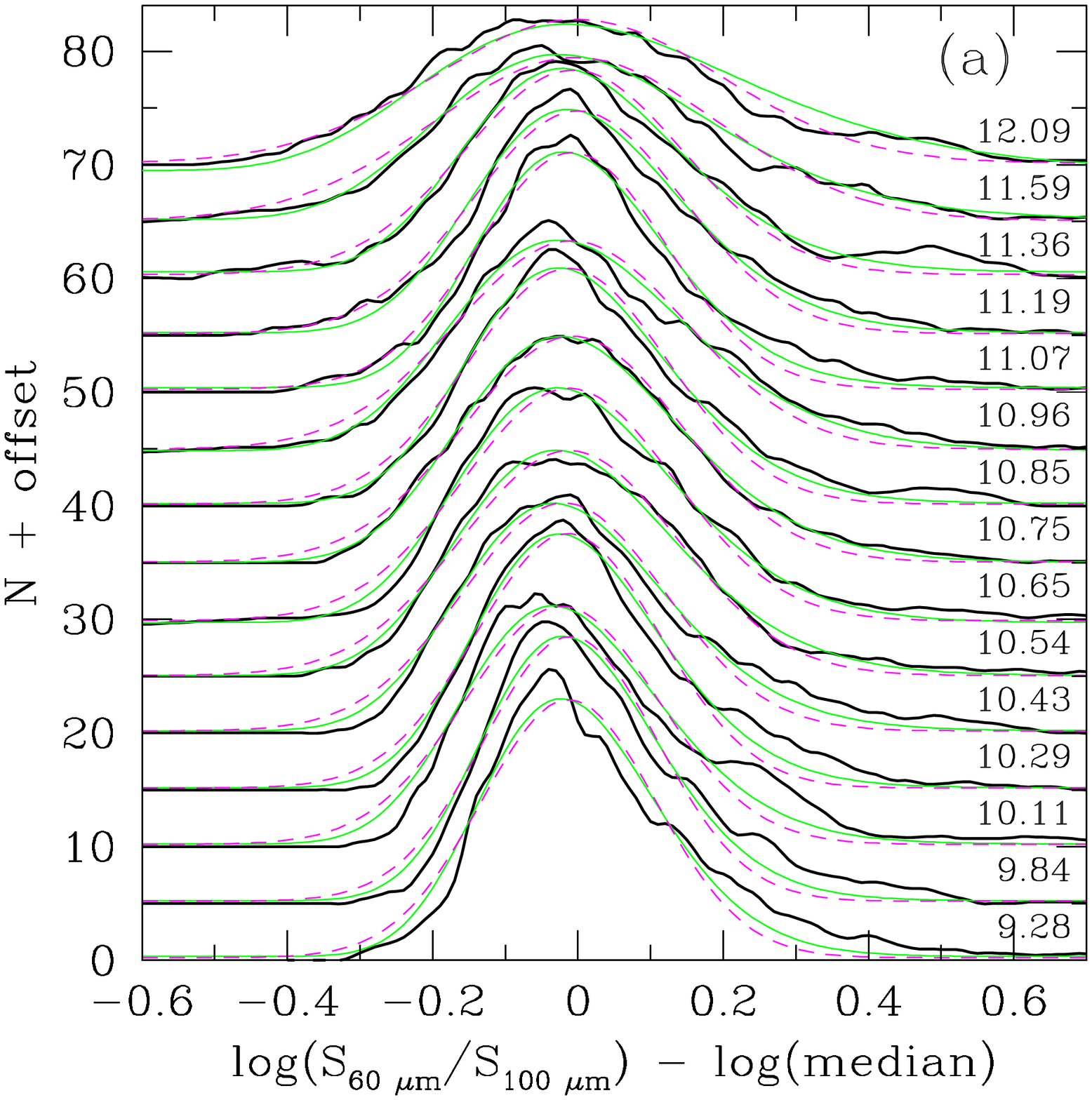,angle=0,width=3.5in}
\psfig{figure=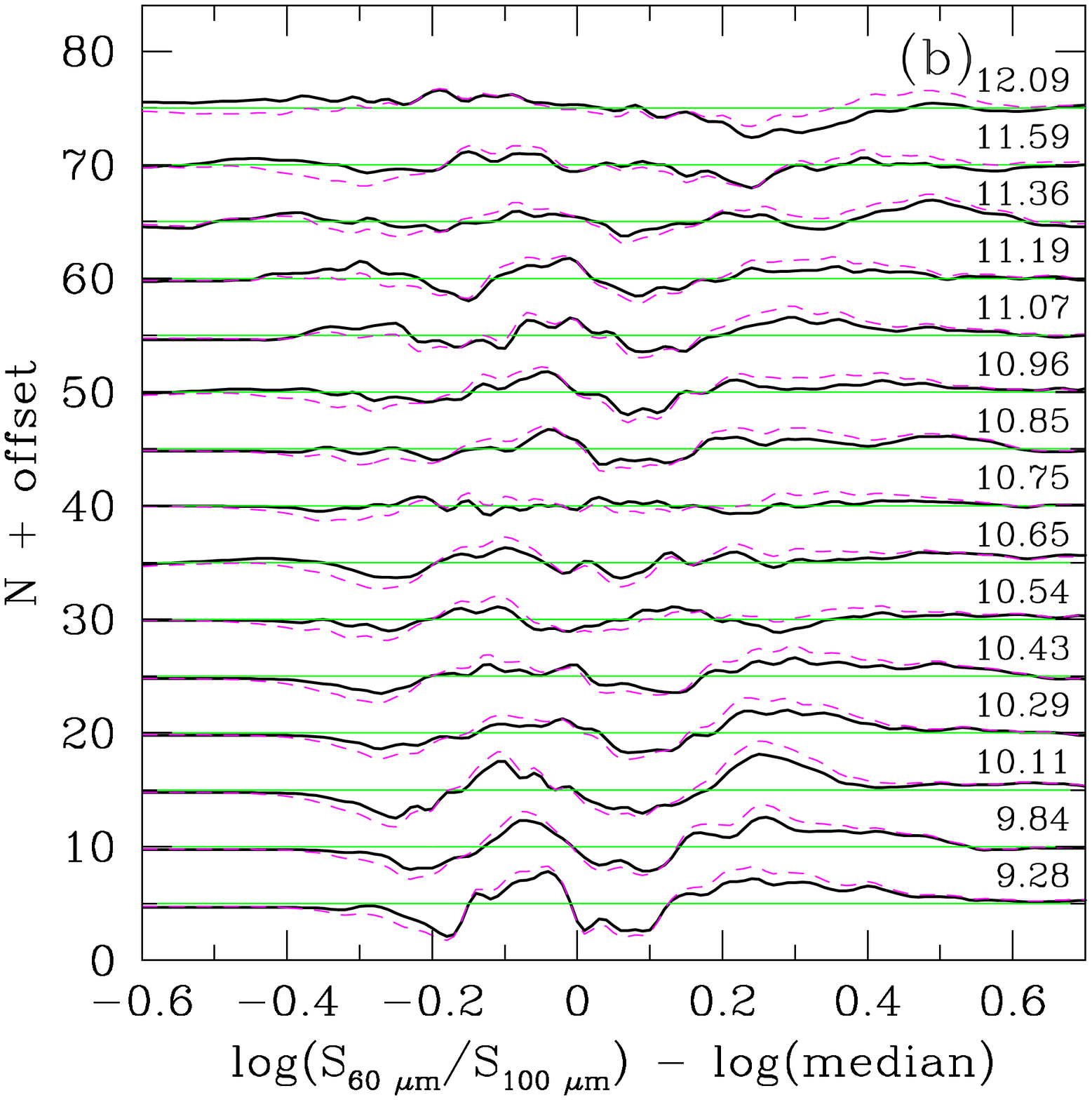,angle=0,width=3.5in}
\vspace{6pt}
\figurenum{4}
\caption{
{\bf upper panel:}
The \ratio\ distribution (heavy solid lines)
with log normal fits (light green lines) and linear Gaussian fits
(light dashed magenta lines). Offsets for each luminosity class 
are inserted as in Fig.~\ref{fig3}. 
The median \ratio\ has been removed for comparison
of luminosities. {\bf bottom panel:} The subtracted residuals
are shown for both log normal (heavy line) and linear gaussian fits
(dashed light line).
Note that an improved fit, in the sense of reduced $\chi^2$ cannot be
obtained by a first order term of the form $x$*G. A higher order term
is required to obtain a better fit, which is not justified
for the present data set.
}
\label{fig4}
\addtolength{\baselineskip}{10pt}
\end{inlinefigure}

The results of our fitting are shown in Fig.~\ref{fig4}a, 
with each luminosity bin of the color histogram offset 
for clarity. The median has been removed from each histogram in order to 
compare directly the form of the distributions at various values of \ltir.
Fits of log normal distribution (light green line) and linear gaussian
(light dashed magenta line) are shown. 
The residuals are plotted in Fig.~\ref{fig4}b, 
emphasizing the superior fit of the log normal, as well as the 
skewed nature of the
distribution. Heavy lines show the residuals from a log gaussian fit, while
the light dashed magenta lines show the residuals from a linear gaussian fit.
While the skew is an apparently large and systematic effect, 
(it appears as a double-S, integral shape) the most significant deviations
occur in the wings of the distribution, where our error bars
are larger than the skew due to the rapidly diminishing number of sources
in each histogram bin.
For completeness we attempted to fit the skewed component
with a multiplicative factor $x\times \exp [-z^2/2]$, but that did not
improve the $\chi^2$.
The systematic skew to the distribution
therefore cannot be removed without a term which is too high an order to justify
from the statistics of the sample. 
We therefore use the log-normal distribution fit as the 
best representation of the 1.2\,Jy IRAS galaxy sample.
Note also that the residual skew is negligible near the intermediate luminosity
bins of the population (the best sampled region) and becomes more pronounced
towards the luminosity extremes.
The skew may therefore partially be a result of including broader luminosity
ranges in the distribution to retain equal number bins.

The distribution over \ratio\ color can then be expressed as

$${\cal G}({\cal C}) = exp(-1/2 \times [({\cal C} - {\cal C_0}) / \sigma_{\cal C}]^2),$$
$$ {\rm where}\ \sigma_{\cal C} = 0.065,\, {\cal C} = \log{\cal R}{(60/100)},$$
$${\cal C_0} = {\cal C_*} \times (1 + {{\cal L_*}\over{L_{\rm TIR}}})^{-\delta} \times (1 + {{L_{\rm TIR}}\over{\cal L_*}})^{\gamma}$$ 
$${\rm with,}\ \gamma = 0.16,\, \delta = 0.02,\, {\cal C_*} = 0.45,\,
        {\cal L_*} = 5.0\times10^{10} L_\odot$$

As the residual skew is close to symmetrical about the gaussian fit
peak, we find that our fit to the distribution describes nearly the
same function as the median and first/last quartile 
expression (a log-normal) presented in the previous section.
The formal \bivar\ is then expressed as follows, where we 
have derived the TIR luminosity function directly from the 1.2Jy
catalog as described in the previous section,
$$ \Phi({\cal L}, {\cal C})\ {\rm d}{\cal L}\, {\rm d}{\cal C} =
  \Phi_1({\cal L}) \times \Phi_2({\cal C})\ {\rm d}{\cal L}\, {\rm d}{\cal C}$$
  $$= \rho_* \times {{\cal L}\over{\cal L_*}}^{(1-\alpha)} \times 
	(1 + {{\cal L}\over{\cal L_*}})^{-\beta} 
	\times \exp[-1/2({{{\cal C} - {\cal C_0}}\over{\sigma_{\cal C}}})^2]
{\rm d}{\cal L}\, {\rm d}{\cal C},$$
$$ {\rm with}\ {\cal L} = {\rm L_{TIR}},\ 
	{\rho_*}=4.34\,Mpc^{-3}\,L_\odot^{-1},\ {\cal L_*}=10^{10.7}\,L_\odot,\ 
{\alpha} = 1.55,\ {\beta} = 2.10,$$
$$ {\cal C} = \log (S_{60 \mu m}/S_{100 \mu m}),\ 
	{\cal C_0}\ {\rm as\ above},\ {\sigma_{\cal C}} = 0.065.$$

%
%
\begin{inlinefigure}
\psfig{figure=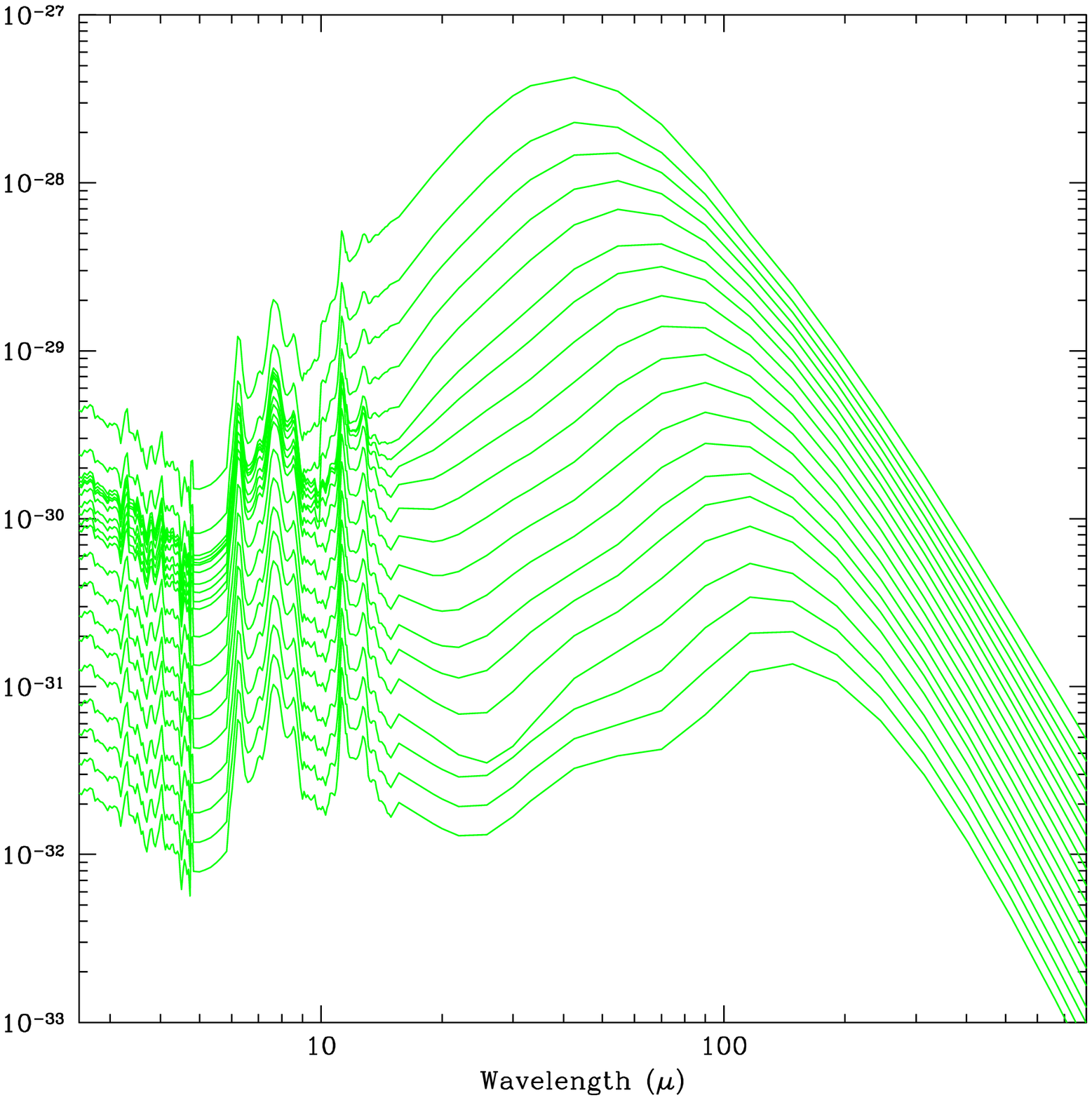,angle=0,width=3.5in}
\vspace{6pt}
\figurenum{5}
\caption{The range of local SEDs used in our model
from the catalog of Dale et al.\ (2001, 2002). The flux units are
ergs/s/cm$^2$/Hz normalized to the median luminosity of a $z=0.1$
galaxy in our model.
}
\label{fig5}
\addtolength{\baselineskip}{10pt}
\end{inlinefigure}

Our analysis has suggested that the ${\cal L}-{\cal C}$ distribution is best
described with a dual power-law in ${\cal L}$ with a break luminosity
$5.0\times10^{10} L_\odot$, a faint end slope of $\delta = 0.02$ and
a bright end slope of $\gamma = 0.16$.
We offer two possible physical interpretations of this result.

Firstly, this may be the point of transition from cirrus-dominated
luminosity to active star formation (i.e.\ high-density photo-dissociation
regions, etc.) dominating the luminosity.
The different power laws then arise because in the former case the luminosity 
increases mostly by making the emitting dust mass larger,
whereas for active star formation, the heating drives the luminosity.

Secondly, this may be the point of transition from where
the 60\mum\ band still
has fluctuating grain emission contributing, to
having the 60\mum\ band dominated by large dust grains.
Once large dust grains dominate the 60\mum\ flux, 
the \ratio\ ratio begins to look like a black-body ratio, leading.
naturally to the broken power-law relation. 
However in this case, the high luminosity, steep portion of the relation
should scale with dust temperature like ${\cal L}\propto T^5$. 
Since this is not observed, this second explanation cannot dominate the
observed relation.

\section{consequences of the ${\cal L}-{\cal C}$ relation for evolution 
models}\label{evolve} 

Several authors have recently modeled the evolution of dusty galaxies
using pure, or nearly pure, luminosity evolution,
reproducing the observed counts and backgrounds
at IR through sub-mm wavelengths
(e.g.~Blain et al.~1999a,b; Malkan \& Stecker 2000;
Rowan-Robinson 2001; Chary \& Elbaz 2001; Chapman et al.~2002b; 
Franceschini et al.~2001).
None of these models include a bi-variate
luminosity function (LF), 
and at best map the dust temperature or spectral shape monotonically
to the luminosity.

%
%
\begin{figure*}[htb]
\centerline{
\psfig{figure=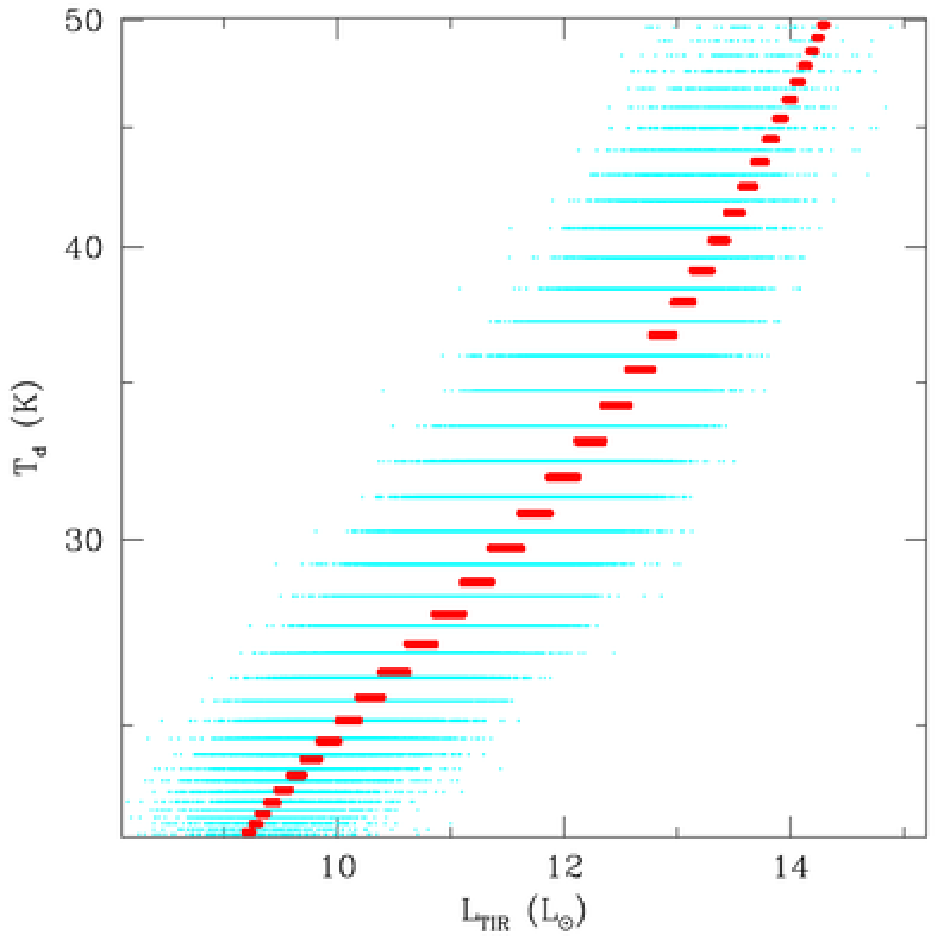,angle=0,width=5.5in}}
\figurenum{6}
\caption{A plot of the  L-T$_{\rm d}$ plane showing points drawn from
our Monte Carlo sampling of the evolving bi-variate LF,
and depicting all objects in the simulated volume of the survey
regardless of their detectability by various instruments.
For comparison we show a representation of the same evolving model 
with a very narrow range of colors for each
luminosity (bold, offset points).
The absissa has been mapped from \ratio\ to dust temperature
(T$_{\rm d}$) using the single temperature greybody which provides the
best fit to the SED template. The quantized appearance of the points
in T$_{\rm d}$ results from the individual 64 SED classes considered.
The LF with narrow range of IR color closely approximates a
single variable LF whereby luminosity maps monotonically to T$_{\rm d}$ or
IR color.
}
\label{fig6}
\addtolength{\baselineskip}{10pt}
\end{figure*}

%
%
\begin{figure*}[htb]
\centerline{
\psfig{figure=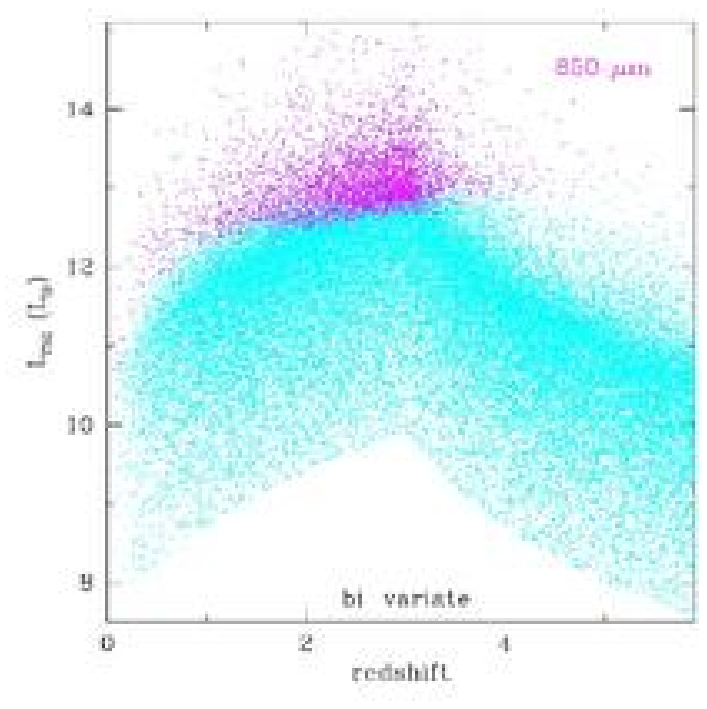,angle=0,width=3.5in}
\psfig{figure=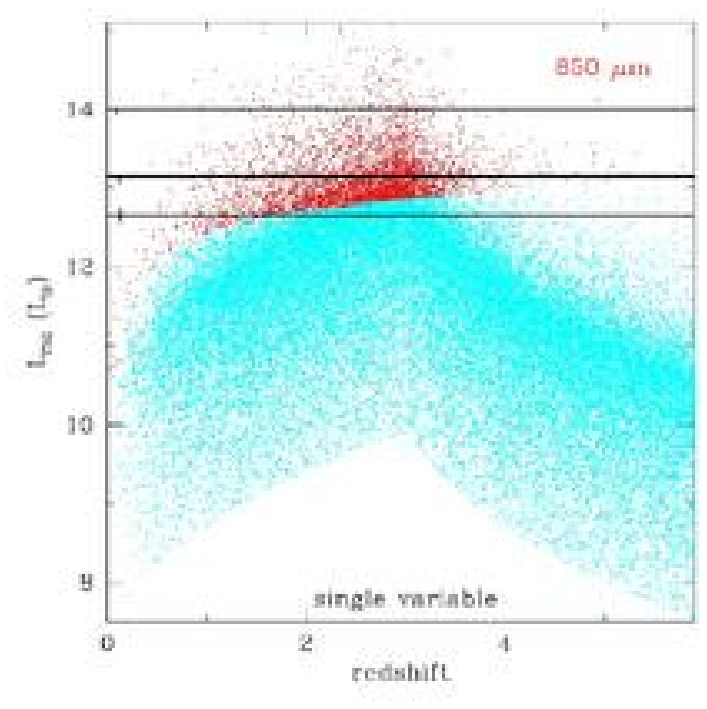,angle=0,width=3.5in}}
\centerline{
\psfig{figure=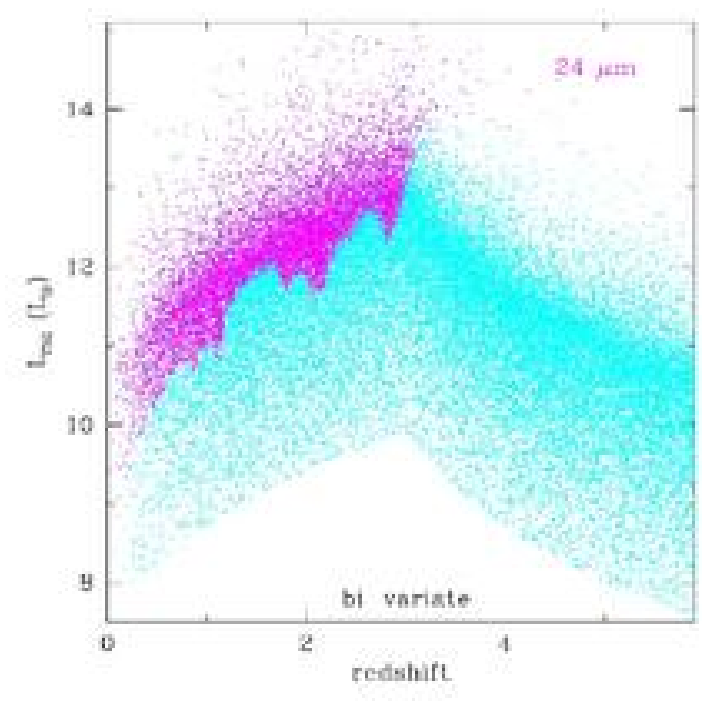,angle=0,width=3.5in}
\psfig{figure=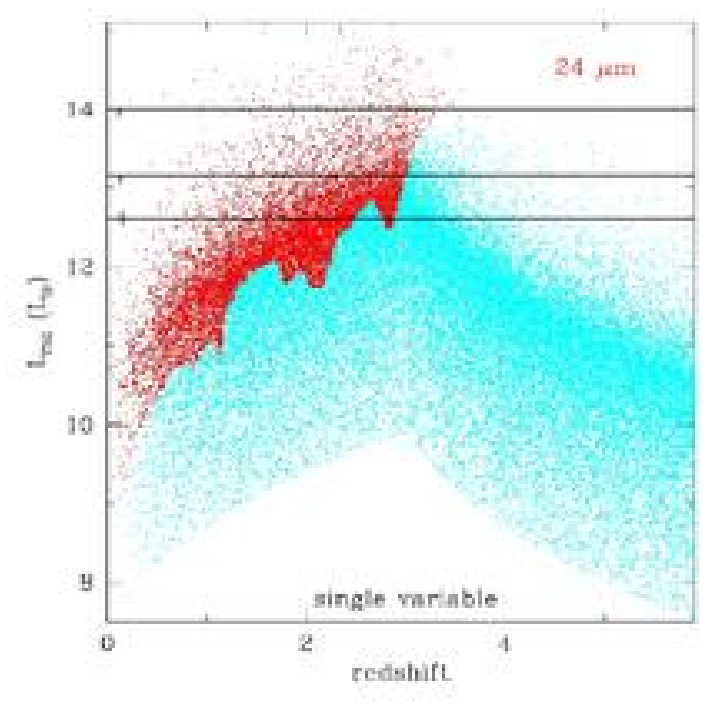,angle=0,width=3.5in}}
\figurenum{7}
\caption{
Comparison of our Monte Carlo representation of the evolving
\bivar\ to an evolving model LF with a one-to-one mapping of luminosity 
to color.
The lighter symbols represent only luminosity and thus follow
identical distributions in both cases. The difference between models is 
apparent when flux limited surveys are defined on the model points.
We overlay flux limited samplings of the model points 
at 24\mum\ and 850\mum\ with darker symbols.
Structure in the 24\mum\ survey is a result of PAH bands (rest $\sim$10\mum)
being redshifted through the SIRTF 24\mum\ filter.
In the single variable LF
model, luminosities map uniquely to fluxes and lines can be drawn
which reflect fixed temperatures, shown at 50\,K, 38\,K and 35\,K, 
top to bottom.
}
\label{fig7}
\addtolength{\baselineskip}{10pt}
\end{figure*}

The spectral shape of the far-infrared background (FIRB) detected by
{\it DIRBE} at 140 and 240\,\mum\ (Puget et al.\ 1996; Fixsen et
al.\ 1998) indicates a peak at $\sim200$\,\mum.  
However, the width of the peak suggests that galaxies
over a large range in redshifts and/or dust temperature contribute to
the FIRB.
There is a degeneracy of dust temperature with redshift (Blain 1999),
since both translating a source to higher redshift and decreasing the dust
temperature will shift the SED to lower frequency. This 
suggests an urgency to explore the effect of the bi-variate LF on 
evolutionary models.

\subsection{A color added evolution model}

Our goal is to  
understand the key differences in \bivar\ from the single variable
$\Phi({\cal L})$. 
We turn to Monte Carlo simulations of a
pure luminosity evolution paradigm, similar to the models of the
above authors.
In companion papers (Lewis et al.\ 2003, S.\ Chapman in preparation), 
we demonstrate that
our model is able to simultaneously fit the farIR background and 
the counts at various wavelength bands. 


%
%
\begin{figure*}[htb]
\centerline{ 
\psfig{figure=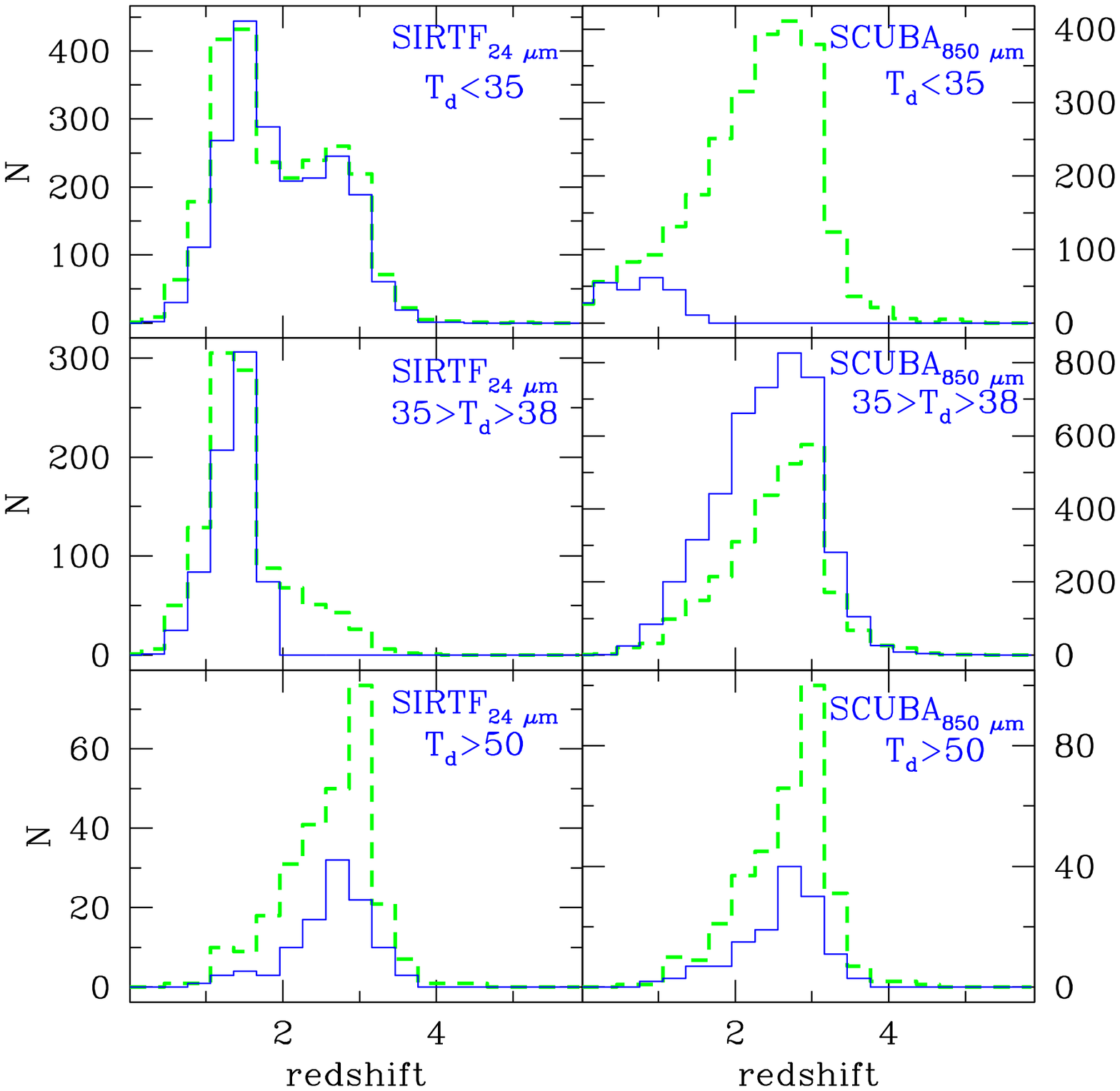,angle=0,width=5.5in}}
\figurenum{8}
\caption{
Histograms of sources with redshift are plotted for the cold, medium
and hot range
in IR color (again mapped to the equivalent single component dust temperature).
Heavy dashed lines represent the evolving \bivar\ distribution, while lighter
solid lines represent the single variable LF.
An excess of hotter sources by a factor $>$2 is revealed for the
\bivar\ model where all sources are luminous enough to be detected regardless
of the observing band. 
For the medium and cold temperature slices, the effective luminosity ranges
lie near the sensitivity limits of the survey which additionally affect
the histograms.
}
\label{fig8} 
\addtolength{\baselineskip}{10pt}
\end{figure*}

We evolve the local FIR
luminosity function (LF) using $\Phi (L,\nu) = \Phi_0 (L/g(z),\nu_0 (1+z))$.
Our evolution function follows a power law in redshift, $g(z) = (1+z)^{4}$,
out to
$z\,{=}\,2.6$. Beyond $z\,{=}\,2.6$
the function drops again as $g(z) = (1+z)^{-4}$ to avoid over-predicting the
farIR background. This power-law index is
chosen based on evolutionary models fit to both optical and sub-mm wavelength
data (Blain et al.~1999a,b). The evolution to $z=2.6$ was chosen as this
provided the best fit to the joint radio/sub-mm sample of galaxies 
(Chapman, Lewis \& Helou~2002, Lewis et al.\ 2003) -- currently the best constraint on high-$z$
farIR galaxy evolution (Blain et al.\ 1999a).
The redshift $z=2.6$ is also close to the 
consensus of the median redshift for SCUBA sources
suggested for example by photometric redshifts in Ivison et al.~(2002).
In order to match the pure luminosity evolution, we truncate the evolved
functions at low luminosity and at both extremes of color, 
so that they integrate to identical numbers of sources per
comoving volume.

One additional assumption in this model is
that the bi-variate distribution, \bivar, evolves
in ${\cal L}$ as a function of redshift, while the relation as observed 
locally between ${\cal L}$ and ${\cal C}$ holds at all redshifts. 
The width
of the distribution in ${\cal C}$ follows our measured relation from section 3,
which remains close to constant in $\log$(\ratio).
This implies that the characteristic
dust temperature of higher redshift sources will be greater because of the
rise in ${\cal L}$. 

We note however, that while the no evolution hypothesis for ${\cal C}$ is
the simplest, it is not necessarily the obvious choice physically.
Higher redshift sources may have distinctly different dust properties
than local analogs. For instance,
metallicities will likely be lower, there may be less
dust to heat, and dust may be less centrally concentrated.
This form of evolution might lead to the
\ratio\ distribution being biased to cooler IR colors with increasing redshift.
In the absence of any strong constraint from the counts or background
(Lewis et al.\ 2003, S.\ Chapman in preparation),
this assumption needs to be tested by future
observational data sets.
However, we shall examine in the following section the preliminary
evidence in support of this simple hypothesis.

We then incorporate this hypothesis into our Monte Carlo models,
drawing \ratio\ values from the evolving \bivar\
to fill our desired survey volume. This model does not include observational
error, but should accurately represent the chance of observing a source with
a given IR color in a particular survey.
The simulation is cut off at 10$^{8}\lsun $ in order to avoid truncation   
effects from the fixed comoving number density in the luminosity functions.

In order to tie our evolving ${\cal L}$(TIR) luminosity function 
to observable quantities,
we map each galaxy with a given \ltir\ and \ratio\ value to
a spectral energy distribution (SED) shape. Normalization of the SEDs is 
accomplished by integrating the SED over the 3--1100\mum\ range and 
scaling to the adopted \ltir.
Spectral templates are taken from the Dale et al.\ (2001, 2002) catalog,
divided into 64 classes from \ratio=0.29 to 1.64,
corresponding roughly to single component dust temperature models of
19\,K to 56\,K. Representative SEDs from our catalog are shown in 
Fig.~7. 
The global galaxy SEDs adopted here use a power-law distribution of dust over
heating intensity ${\cal U}$ in order to reproduce the range of 
photometric and spectroscopic properties observed by IRAS and ISO for
galaxies in the local Universe.
Dale \& Helou (2002) specifically investigate the dust emissivity
of the local IRAS galaxies, using the sub-mm data presented in Dunne et al.\ (2000). They find the dust emissivity should vary with environment with
$\beta = 2.5 - 0.4\log {\cal U}$, where ${\cal U}$ 
is the local radiation field.

%

\subsection{Model Results and Implications}

To underscore the effect of not including the \bivar\ distribution
in the redshift evolution,
we show in Figs.~6--8 the output from our luminosity evolution models,
with and without the broad local color distribution in the LF. 

In Fig.~6, we plot the ${\cal L}$-T$_{\rm d}$ plane showing points drawn from
our Monte Carlo sampling of the evolving bi-variate LF.
We compare directly the bi-variate
\bivar\ to an equivalent $\Phi({\cal L})$, 
with a very narrow range of colors for each 
luminosity (bold, offset points).
The ordinate has been mapped from \ratio\ to dust temperature
(T$_{\rm d}$) using the single temperature greybody which provides the
best fit to the SED template. The quantized appearance of the points
in T$_{\rm d}$ results from the individual 64 template SED classes considered.
The LF with narrow range of IR color closely approximates a
single variable LF whereby luminosity maps monotonically to T$_{\rm d}$ or 
IR color.
Fig.~6 provides an overview of \ratio\ as a function of ${\cal L}$, simply 
extrapolating the local ${\cal L} - {\cal C}$ relation to the required 
luminosities -- our baseline assumption in this model.
The distribution is symmetric in $\log$(T$_{\rm d}$) about each luminosity.
As the \bivar\ distribution remains identical to our locally characterized
form (\S~3) for all redshifts, any deviations from the
median local relation reflect only the
scatter in the relatively small numbers of luminous galaxies in this
simulated survey volume.

In Fig.~7, we show the \ltir\ distribution as a function of redshift.
The left panels show the \bivar\ model, while the right panels show the
single variable LF model. Until we apply a flux limit to the 
figure (darker points), there is no difference between the visualizations
since they have the same luminosity evolution formalism.
Fig.~7 is our Monte Carlo representation of the evolving LF; vertical slices
reveal the dual power law $\Phi({\cal L})$ at each redshift.
Note that 
with comoving number density, in the absence of luminosity evolution,
a local physical density of N objects per Mpc$^3$ corresponds to a physical
density of N * $(1+z)^3$ at redshift $z$.
In this type of model the evolution is not necessarily meant to imply that
ULIRGs fade into less luminous LIRGs with time. Instead, it provides
a picture where, for instance, the local counterparts with
volume densities similar to ULIRGs at $z\sim1$
have 10$^{11.5}$\,L$_\odot$
at the present-day.
This form of evolution is then {\it not} describing the
internal physics of the IR luminous population; high-$z$
ULIRGs do not evolve into similar local ULIRGs, but instead likely
form ellipticals (e.g., Sanders \& Mirabel 1996; Tacconi et al.\ 2002), 
fading away from the LF.

When flux limited surveys are considered in the context of Fig.~7 
(850\mum\ -- top, 24\mum\ -- bottom),
differences in the two models become manifest. 
In the case of the single variable distribution, each \ltir\ point maps 
uniquely to a flux for a given wavelength. However in the bi-variate LF, 
each \ltir\ point corresponds to a probability distribution of fluxes
corresponding to the log-normal distribution in \ratio\ and the associated
range of SED templates that can be tied to the \ltir\ value.
The sensitivity limits in the 850\mum\ and 24\mum\ bands
are shown for characteristic survey depths with SCUBA and SIRTF
(2\,mJy r.m.s., and 0.1\,mJy r.m.s., respectively). The 
structure in the 24\mum\ survey is a result of PAH bands (rest $\sim$10\mum)
being redshifted through the SIRTF 24\mum\ filter.
At 850\mum, the K-correction from the rising grey-body dust SED 
results in a flat Luminosity-Flux relation for redshifts $z$ \cge 1.

The effect is subtle in Fig.~7, as both the luminosity function and
the color distribution are scattering the observed fluxes,
largely canceling dramatic differences in the effective luminosity limits 
probed with redshift. 
However, the most important difference between the bi-variate and luminosity-only 
models is apparent in Fig.~7: in the simpler  $\Phi({\cal L})$ model,
the flux limit for a given wavelength translates at each redshift
into a transition range of luminosities, within which galaxies are or are not
detected depending on their color.
We draw lines on the single variable model for ${\cal L}$ corresponding
to T$_{\rm d}$ of 35\,K, 38\,K, and 50\,K, using the mapping illustrated
in Fig.~6.


The implications of this model difference are therefore much more apparent in a
direct analysis of T$_{\rm d}$, as demonstrated in Fig.~8. 
Histograms of sources versus redshift are plotted for the 
850\mum\ and 24\mum\ bands, comparing the detectability of
hotter and cooler IR colors 
(again mapped to the equivalent single component dust temperature).
Heavy dashed lines represent the evolving \bivar\ distribution, while lighter
lines represent our approximation to a single variable LF.

We first consider the model differences for a SCUBA\,850\mum\ survey
in Fig.~8. 
A cut for detectable sources colder than T$_{\rm d}<$35\,K (shown in Fig.~7)
lies along the sensitivity limit of SCUBA. This results in the most
dramatic difference between the models, since in the bi-variate model,
the steep luminosity function
scatters many more cold, under-luminous sources into the sample than the
corresponding loss of hotter, higher luminosity sources.
The selection along the cold side of the grey-body peak for SCUBA
therefore produces a substantial population of cold sources
which are strikingly absent when only the single variable LF is included in the
evolution. 
This difference results in
an important prediction of the existence of cold and luminous galaxies
at moderate to high redshifts -- a prediction which has been verified
through the detection of ULIRGs with T$_{\rm d}<30$\,K (Chapman et al.\ 2002a).
The opposite effect is shown in the middle panel of Fig.~8 for the range
35$<$T$_{\rm d}<$38\,K, producing a band in luminosity lying just above
the sensitivity limit of our SCUBA survey in Fig.~7. The bi-variate model
now scatters the warmer sources below our detection limit, resulting in a dearth
of sources relative to the single variable LF.
For the hottest sources (T$_{\rm d}>50$\,K), the luminosities are large
enough (Fig.~7) that the sensitivity limit of the survey has no bearing on the
observed sources, and the comparison of models simply reflects the steep
luminosity function which asymmetrically scatters more low luminosity hot
sources into the temperature cut than luminous cold sources which fall beneath
the cut.
The bi-variate \bivar\ model thus predicts almost twice as many
warm galaxy detections, even though the underlying luminosity distribution
is the same as in the $\Phi({\cal L})$ model.

For surveys
selecting sources along the hot dust side of the grey-body peak, that being
the case for all the accessible wavelengths of SIRTF except 170\mum, 
colder luminous sources will
be missed and hotter low luminosity sources will be preferentially detected.
This is exactly in the opposite sense to the situation 
for sub-mm selected surveys (such as with SCUBA), preferentially detecting
colder sources at a given luminosity (Eales et al.\ 1999, Blain et al.\ 2002). 

The sensitivity limit of SIRTF is adversely affected by the steep Wien
slope of the far--mid-infrared SED,
making more distant sources difficult to detect and
resulting in the steep sensitivity curve in Fig.~7.
The result in Fig.~8 is that the histograms near the sensitivity limit of SIRTF
will be dominated by the large numbers of sources lying both above and
below any fixed temperature cut. 
The histograms in the cold and warm 24\mum\ panels therefore continue
to show the asymmetric {\it shuffling} of sources due to the steep luminosity
function, but only as a small perturbation on the large number of sources
which are unaffected by the sensitivity limit.
However, for hotter dust temperatures, an excess of a factor 
greater than two of sources is predicted by the
\bivar\ model. 
This excess occurs for the SIRTF 24\mum\ band for the same reason discussed
above for SCUBA.
All sources are luminous enough to be detected regardless of their temperature,
and there are far more lower luminosity sources which
are boosted by the bi-variate LF 
into the T$_{\rm d}>50$\,K bin than vice versa.

Consideration of the model differences in Fig.~8 also addresses 
the question of the SCUBA/SIRTF galaxy overlap.
As the SCUBA and SIRTF observing windows lie on opposite sides of the
dust spectral energy peak, the direct overlap of the populations is a strong
function of the dominant dust temperature. 
In particular, a cold (T$_{\rm d}<30$\,K) SCUBA population may be difficult 
to detect even in the deepest SIRTF exposures, as revealed in the
high redshift excess of cold SCUBA sources compared with the SIRTF population. 
Scrutiny of Fig.~7 along the survey flux limits 
reveals that at $z$\cle 3, the individual
SCUBA galaxies detected in either model are mostly 
detected by SIRTF at 24\mum\ at this survey depth. 
However, deeper sub-mm surveys by instruments like the {\it
Large Millimeter Telescope} (Hughes et al.\ 2001) may
uncover fainter cold galaxies at lower redshifts which would require very 
deep SIRTF exposures to detect.
By contrast, most or all of the
SCUBA sources should be detected by SIRTF over all redshifts 
when the characteristic dust
temperatures are higher, as suggested by the similar N($z$) in the lower
panels of Fig.~8.
 
Figs.~7\&8 are also suggestive of the effect of the evolutionary
form on the detectability of galaxies at different wavelenghs.
The properties of the sources detectable at high redshift are a sensitive
function of the evolutionary form adopted -- in our scenario, lower redshift
peaks in the evolution function would result in an appreciably smaller 
fraction of high redshift galaxies detected (effectively generating 
differences in the redshift distributions).
For example, a direct application of our \bivar\ model to the SCUBA population
(Chapman, Lewis \& Helou 2002, 
Lewis et al.\ 2003) has suggested a best fit peak at a
lower redshift of $z\sim2.5$. 
However, this is extremely sensitive to 
any evolution in the IR color distribution (but see \S~5 to follow). 
We defer detailed fitting of our model to the
available high redshift counts and FIRB to a companion paper (Lewis et al.\
2003). 
Ultimately, redshift surveys of SIRTF and SCUBA galaxies will be 
required to test the detailed form of the evolution function
(see Chapman et al.~2003a for preliminary results on the SCUBA galaxy
redshift distribution).

%
%
\begin{figure*}[htb]
\centerline{
\psfig{figure=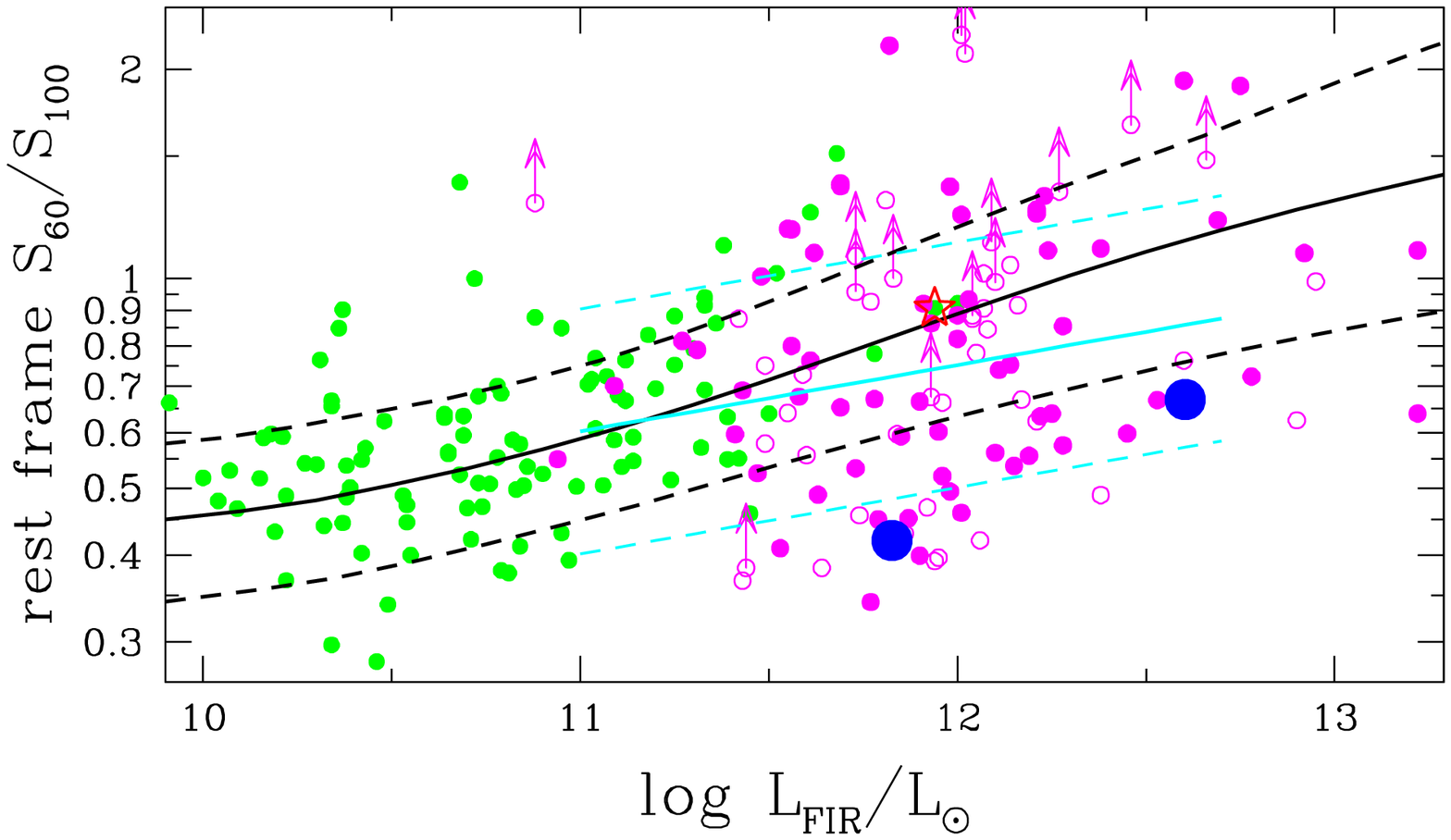,angle=0,width=6.3in}}
\centerline{
\psfig{figure=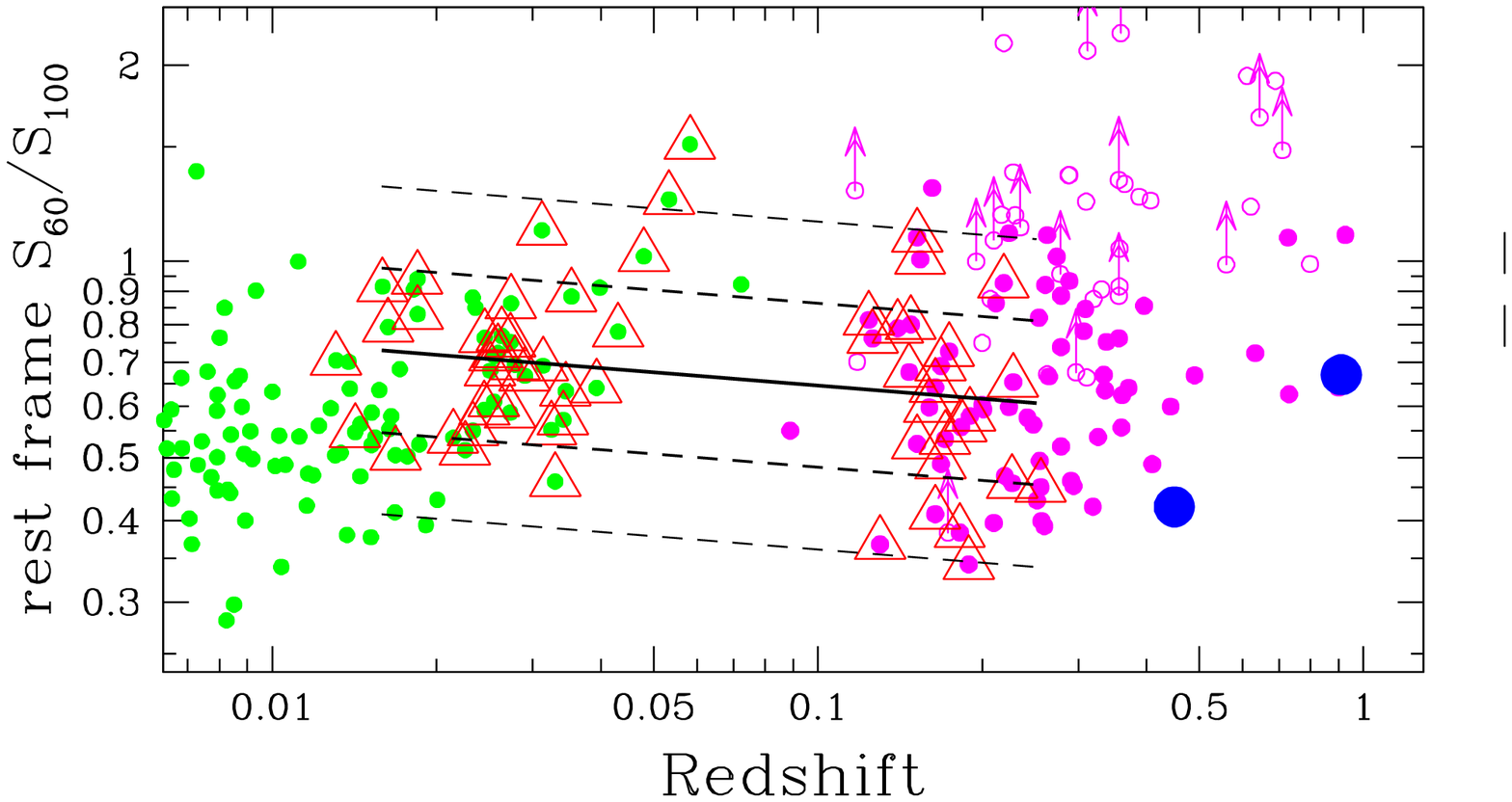,angle=0,width=6.3in}}
\figurenum{9}
\caption{
{\bf upper panel:}
The dependence of rest frame \ratio\                     
(roughly dust temperature) on L$_{\rm FIR}$ is shown.
LIRGs and ULIRGs from the Stanford et al.~(2000) FIRST/60\mum\ selected sample,
with redshifts from 0.1 to 0.9 (magenta circles), appear as a consistent
extension to local IRAS-BGS
galaxies (green circles) 
Many Stanford et al.\ sources are detected at 60 and 100\mum\
(solid circles), although some have only tentative 100\mum\ measurements when
analysed in XSCANPI (open circles);
those without 100\mum\ detections at all are shown as limits.
The fit to the Stanford et al.\ sources between 11.5$<\log$\lfir$<$12.5
is shown with lighter solid and dashed lines, for the $\pm$1$\sigma$ envelope.
Two FIRBACK 170\mum\ selected galaxies
from Chapman et al.~(2002a) are shown as large circles.
Our best fit correlation from section~3 and 1$\sigma$
deviations are overlaid as the thick solid and dashed lines.  
Arp220, the ULIRG typically used for comparison
with high$-z$ luminous galaxies, is identified with a star.
{\bf lower panel:}
The \ratio\ ratio is shown as a function of redshift for the same sources.
Overlapping luminosity regions can be compared directly.
Triangles have $\log$\lfir$\sim$11.5$\pm$0.5. 
Fits to the population overlapping in luminosity
at \ltir=3$\times10^{11}\,L_\odot$ 
signalled by oversize triangles, 
is shown with a solid line, along with  1$\sigma$,2$\sigma$ distributions
as dashed lines.
The slopes of the fit line is not significantly different from zero.
}
\label{fig9}
\addtolength{\baselineskip}{10pt}
\end{figure*}

\section{Comparison with data out to higher redshifts}

Our comparison of evolving a single variable LF versus a bi-variate 
\bivar\ (Figs.~6--8) clearly demonstrates that an over simplified 
representation of the galaxy population can miss large classes of
detectable sources at high redshifts.
It is then crucial to understand whether the IR color distribution itself
has evolved with redshift.

While much attention has been paid to the luminosity evolution
of ULIRGs (e.g.\ Blain et al.~1999a,b; 
Chary \& Elbaz 2001; Malkan \& Stecker 2001; Rowan-Robinson 2001;
Chapman et al.~2002b; Ivison et al.\ 2002),
there has been no study of the \ratio\ evolution.
The simplest scenario, which we considered above, is that the \ratio\ 
correlation with
L$_{TIR}$ continues to higher luminosities, extrapolating the
expressions derived in this work for 
the local \ratio\ distribution. 

Existing data sets are largely inadequate for characterization of the 
high-$z$ IR color distribution. ISOPHOT and SCUBA galaxies do not generally
have accurate redshifts or even detections at sufficient wavelengths to measure
the rest frame \ratio.
However, a survey of moderate redshift, luminous IRAS galaxies
(Stanford et al.\ 2000) provides an initial foray into the higher
redshift \ratio\ distribution (for a complementary analysis of this
population, see also Blain, Barnard \& Chapman 2003).
This section will focus on the 
Stanford et al.\ IRAS galaxies, as well as 
the microjansky radio sources (Chapman et al.\ 2003b) to assess our 
\bivar\ characterization. 

\vskip3cm

\subsection{60\mum-selected LIRGs/ULIRGs at $0.1<z<0.9$}
A recent sample of distant LIRGs and ULIRGs from Stanford et 
al.~(2000) has an effective sensitivity only 2-3 times higher than 
that of the deepest ISOPHOT 170\mum\ surveys, 
and we can use this sample to study the higher redshift \bivar.
The Stanford et al.\
sample was selected from a positional cross--correlation of the
$IRAS$ Faint Source Catalog with the FIRST database.  Objects from
this set were selected for spectroscopy by virtue of following the
well-known star-forming galaxy correlation between 1.4\,GHz and 60\mum\
 flux, and by being optically faint on the POSS.  Optical
identification and spectroscopy were obtained for 108 targets at the
Lick Observatory 3~m telescope.  Most objects show spectra typical of
starburst galaxies, and do not show the high ionization lines of
active galactic nuclei.  The redshift distribution covers $0.1 < z <
0.9$, with 13 objects at $z > 0.5$ and an average redshift of $\bar{z}
= 0.31$.

\subsubsection{The dependence of \ratio\ on ${\cal L}$}

We first address the issue of the dependence of width in
${\cal C}$ (= \ratio) on the luminosity. Fig.~9a 
shows the \ratio\ versus {\cal L} distribution of the Stanford et al.\ sample 
relative to our fit ${\cal C}$ relation from section~3, overlaying
the local BGS IRAS galaxies
observed with SCUBA by Dunne et al.~(2000) for reference. 
The rest frame \ratio\ has been calculated by 
fitting the Dale et al.~(2001, 2002)
template SEDs to the observed frame 60\mum, 100\mum, and radio points, and
extracting the restframe 60\mum, 100\mum\ fluxes.
The moderate$-z$ sources from Stanford et al.~(2000) are not, however, 
all detected
at 100$\mu$m. We have run the IPAC application {\bf XSCANPI} 
over all 103 Stanford sources to derive improved estimates of the
60 and 100\mum\ fluxes,
finding 46 with spurious 100\mum\ measurements. 
The sources without 100\mum\ detections are presented as lower limits
to ${\cal R}(60,100)$,
while those with only marginal 100\mum\ detections are differentiated with
open symbols. The fits to the distribution
are then derived, using only sources with secure 100\mum\ detections.
The fit result is shown as the lighter solid line with the $\pm1\sigma$
lines shown as lighter dashed lines in the upper panel of Fig.~9.
The sources with 100\mum\ limits appear to be uniformly distributed 
amongst the 100\mum\ detected sources, and a large systematic skew is
unlikely.

The median relation between \lfir\ and \ratio\ is similar to that observed 
locally, although it falls to the -1$\sigma$ deviation of the local
relation for the most luminous Stanford sources. However, these most
luminous sources are not very numerous and contribute negligibly to the
fit lines. 
Moreover, the fit did not take into account the lower limits
to \ratio.
The median of this luminous tail of Stanford sources
on their own is in fact more consistent with the local relation.

The width of the Stanford distribution matches the extrapolation
from the 1.2\,Jy envelope at L$_{\rm FIR}$=3.1$\times10^{12}$\,L$_\odot$
(Fig.~9a). However, the width of the Stanford sources exceed the
local distribution by a factor of $\sim$50\% at 
L$_{\rm FIR}$=4.6$\times10^{11}$\,L$_\odot$,
and there is an apparent excess of LIRG/ULIRG class sources 
as cold as the two FIRBACK galaxies discussed in the introduction, 
FB1-40 and FB1-64 at their respective FIR luminosities (see \S~5.1.3). 
As the Stanford et al.\ sources were selected in the 60\mum\ band, a 
direct comparison with our 1.2\,Jy catalog is justified, with the assumption
that the fainter flux limits of the Stanford et al.\ sample simply allows
a probe of the higher luminosity distribution out to moderate redshifts
where evolution is still not likely to be a large effect.
However, even for sources with detected 100\mum\ fluxes, the signal-to-noise
ratio is
not high, and photometric scatter may affect the width of the diagram.
A further complication is that while
the rest-frame \ratio\ has been demonstrated to
correlate well with cold dust temperature using 850\mum\ measurements (Dunne et
al.~2000), the observed \ratio\ becomes less
effective at constraining the cold temperature properties as redshifts
increase beyond $z\sim0.5$.
The majority of the Stanford sources are at $z$ \cle 0.4 and this should
not result in a large effect on the average sample properties. 
Observations at 850$\mu$m will be required to properly assess the 
contribution of the Stanford sources at higher redshifts to the width of the
distribution. 

We therefore conclude that the direct
extrapolation from the local IRAS galaxies is consistent with 
the width of the Stanford et al.\ sources, with a possibility of
an increasing width at lower luminosities in the Stanford sample.
We note, however, that the width of even
the local distribution at the highest luminosities
is highly uncertain, and a change in the slope of the
broadening function could be in agreement to within the errors.

\subsubsection{The dependence of \ratio\ on redshift}

We now address the possible evolution in the (${\cal L},{\cal C}$) relation
with redshift.
In Fig.~9b, the \ratio\ ratio 
is shown as a function of redshift for the same sources.
The overlapping luminosity regions can be compared directly. 
Triangles have $\log$\ \lfir/L$_\odot$$\sim$11.5$\pm$0.5, 
in both IRAS-BGS and Stanford et al.\ galaxies. 
A fit to the overlapping luminosity populations
(signalled by oversize symbols)
is shown with a solid line, along with 1$\sigma$ and 2$\sigma$ envelopes.

The salient point of Fig.~9b is that the Stanford sources do not appear 
significantly different
in the median than the local comparison sample.
The large fraction of Stanford sources with lower limits and uncertain
100\mum\ fluxes may be responsible for the median point lying somewhat lower
than local galaxies.
The current data therefore do not support 
a scenario in which the median \ratio\ in ${\cal C}$ varies 
significantly with redshift out to $z\sim1$. 
The width in ${\cal C}$ of both luminosity bins is similar in Fig.~9b, and
wider than the local distribution. This is consistent with the
broadening of ${\cal C}$ with luminosity, and suggests that the
redshift evolution affects the width in ${\cal C}$ only in so far as higher
redshift sources have characteristically higher {\cal L}.
However, the statistics for
sources with $z>0.5$ are poor. With identified redshifts for increased 
numbers of sub-mm galaxies (see Chapman et al.\ 2003a), 
this issue will be directly addressable. 

\subsubsection{ISOPHOT 170\mum\ galaxies at intermediate redshift}

Recently, Keck spectra and UKIRT high spatial resolution near-IR
imagery for two of the proposed highest redshift sources from the
FIRBACK-N1 170\mum\ survey were obtained (Chapman et al.~2002a).
These authors found that the redshifts of counterparts 
to the 170\,\mum\ sources
confirm that both sources are ULIRGs, but that their redshifts are
significantly lower  than implied by fitting a typical ULIRG SED to
their farIR/sub-mm/radio SEDs.  This indicates that they have cooler
dust temperatures, T$_d\sim30$\,K, than the canonical ULIRG values
(T$_d\sim45$\,K).  
As previous models have failed to predict the existence of such cold and
luminous sources, these two galaxies
therefore confirm the importance of considering the
full \bivar\ in evolutionary models.

In Fig.~9, 
we overplot
these 2 FIRBACK ULIRGs on the local IRAS BGS and the
Stanford LIRG/ULIRG sample;
FN1-40 and FN1-64 are cold objects, lying near the -2$\sigma$
level below the local median relation.
While FB1-40,64 appear as
cold objects relative to the local distribution in \bivar, there is not yet
enough followup to ISOPHOT sources to know
if ${\cal C}$ has substantially evolved out to $z\sim1$.
FN1-40,64 remain at the periphery of the distribution, but are certainly not
such unusual objects when compared to similar redshift ULIRGs from the
Stanford et al.\ sample.

\subsection{The ${\cal L}$--${\cal C}$ relation for microjansky radio sources 
}

In this subsection, we test the lower luminosity 
${\cal L}$--${\cal C}$ relation out to $z\sim0.6$
using a sample of radio galaxies with 20cm fluxes ranging from
30$\mu$Jy to 500$\mu$Jy from Chapman et al.\ (2003b).
These radio sources were observed in the sub-mm,
and all have spectroscopic redshifts ranging from $z$=0.1--0.6.
From this information we can estimate T$_{\rm d}$ and \lfir\ by assuming
the FIR-radio correlation for star forming galaxies
(\markcite{helou85}Helou et al.\ 1985) and a synchrotron
spectrum with an index of $\alpha=-0.8$ (\markcite{richards00}Richards\ 2000).
T$_{\rm d}$ is estimated by taking the 
SED template from Dale \& Helou (2002) which has the appropriate temperature 
to fit the radio and sub-mm points at the fixed redshift of the source.
Note that we are only indirectly testing the ${\cal L}$--${\cal C}$ relation
with these galaxies (we are estimating both ${\cal L}$ and ${\cal C}$
through local empirical correlations), and we cannot deconvolve 
variations in the ${\cal L}$--${\cal C}$ relation from the 
FIR-radio correlation.
 
We then restrict the sample to those sources with \lfir$<10^{11}\lsun$
as the spectroscopic incompleteness introduces a severe bias for more
luminous (and typically optically fainter) galaxies (as discussed in 
Chapman et al.\ 2003b). This also reduces the radio-bright AGN
contribution to the sample, which will have much larger apparent \lfir\ 
calculated in this manner due to the AGN generated radio excess. 

%
%
\begin{inlinefigure}
\psfig{figure=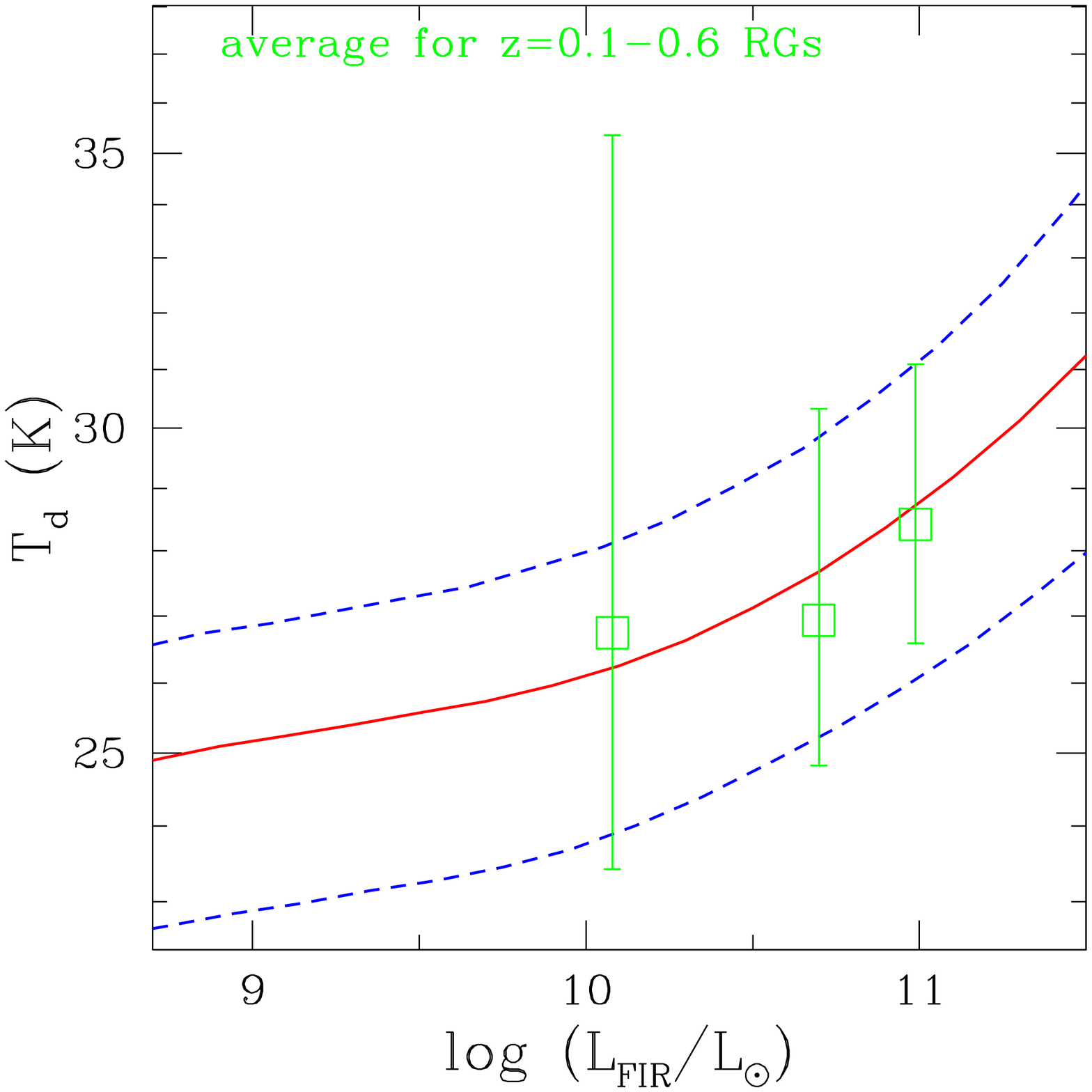,angle=0,width=3.5in}
\figurenum{10}
\caption{
Dust temperature (T$_{\rm d}$) versus the log of FIR luminosity
for radio sources with \lfir$<10^{11}\lsun$
and spectroscopic redshifts. Sources span $z=0.1$--0.6, and have been
divided into three equal number bins. The
T$_{\rm d}$ and FIR values have been calculated from the
redshift and the sub-mm/radio measurements as described in the text.
Our derivation of the
median and interquartile range of local IRAS galaxies from the
1.2\,Jy catalog are shown.
The agreement of the higher redshift sources with the local relation is
remarkable.
}
\label{fig10}
\end{inlinefigure}

In Fig.~10 
we plot T$_{\rm d}$ versus FIR luminosity, along with
our derived local IRAS galaxy relation from \S~3.
The average points in Fig.~10 do not include sources
individually detected in the submillimeter,
and the variance from measurement
errors is expected to be comparable to that from scatter in the
dust properties. Thus our points can only
be used as a characterization of the average ${\cal L}$--${\cal C}$ relation.

Fig.~10 
demonstrates that the \lfir$<10^{11}\lsun$ radio sources, spanning
the redshift range $z=0.1$--0.6 
appear to follow the local IRAS color distribution very well.
This suggests jointly that neither the ${\cal L}$--${\cal C}$ relation
nor the FIR-radio correlation can deviate significantly from
local values for these sources.
~
The analysis provides a direct consistency check of the lower luminosity
tail of local IRAS galaxies with higher redshift star forming galaxies,
of which the microjansky radio population should represent
higher redshift specimens (e.g., Richards 2000).

\section*{Conclusions}

We have analyzed the \ratio\ distribution of local IR-luminous galaxies
finding a best fit, low order analytical expression for the 
bi-variate luminosity function, \bivar.
A log-normal distribution about a dual power law in the median
\ratio\ versus \ltir\ is demonstrated to 
provide the best fit of 
the IRAS population over 4 orders
of magnitude in \ltir.

We then studied the redshift evolution of \bivar, using a luminosity
evolution paradigm.
We demonstrated that while similar luminosity regimes are detectable at
24\mum\ and 850\mum, the dust temperatures 
represented by the detectable sources can differ by factors of $>2$.
For a flux limited survey, the bi-variate \bivar\ 
model predicts about twice as many
warm galaxy detections as the equivalent single variable model, 
even though the underlying luminosity distributions are the same
in the two models.
In a similar manner, recently discovered
populations of cold, luminous galaxies can be predicted naturally within
our model.
Consideration of the model differences also addresses
the question of the SCUBA/SIRTF galaxy overlap,
a strong function of the dominant dust temperature.
A cold (T$_{\rm d}<30$\,K) SCUBA population may be difficult
to detect even in the deepest SIRTF exposures, as revealed in the
high redshift excess of cold SCUBA sources compared with the SIRTF population.

We compare our derived color relation 
with existing data for higher redshift
IR galaxies. General agreement with the local color relation
is found amongst all higher redshift galaxy populations considered.
We therefore find no significant evidence for a variation in the
median \ratio\ ratio as a function of redshift.

We do find tentative evidence for a broadening of 
the IR color distribution at higher ${\cal L}$ from both our local 1.2\,Jy
sample, and the moderate redshift high luminosity sample of Stanford
et al.\ (2000). 
This suggests that 
redshift evolution does not in itself affect the width in ${\cal C}$. 
The baseline assumption 
that the locally measured ${\cal L}-{\cal C}$ relation holds over all redshifts
is therefore a resonable working hypothesis, to be verified with SIRTF data.

\section*{Acknowledgements}
We thank Ian Smail and Andrew Blain for helpful discussions.
An anonymous referee helped to improve the paper.
GFL thanks the Australian Nuclear
Science and Technology Organization (ANSTO) for financial support.
This research has made use of the NASA IPAC Infrared Science Archive, which 
is operated by the Jet Propulsion Laboratory, California Institute of 
Technology, under contract with the 
National Aeronautics and Space Administration.

\end{document}